\documentclass[preprint]{elsarticle}

\usepackage[usenames,dvipsnames]{color}
\usepackage{bm}
\usepackage{graphicx,fancyhdr,lscape,dsfont,amsmath,amssymb} 
\usepackage{color, amsthm}
\usepackage{longtable}
\usepackage{psfrag}
\usepackage{multirow}
\usepackage{stmaryrd}
\usepackage{appendix}
\usepackage{ulem} 
\usepackage{subcaption}

\allowdisplaybreaks

\newcommand{\vect}[1]{\vec{#1}} 
\newcommand{\tensor}[1]{  {\bm {#1}} } 
\newcommand{\fourthorder}[1]{  {\mathds {#1}} } 

\newcommand{\ey}{ {\vect{e}}_{2}}

\newcommand{\trace}[1]{ {\rm tr} \left[ \, {#1} \, \right] }

\newcommand{\determinant}[1]{ {\det} \left[ \, {#1} \, \right] }

\newcommand{\divergence}[1]{ {\rm div} \left[ \, {#1} \, \right] }
\newcommand{\surfacedivergence}[2]{ {\rm div}_{#2} \left[ \, {#1} \, \right] }

\newcommand{\Divergence}[1]{ {\rm Div} \left[ \, {#1} \, \right] }
\newcommand{\surfaceDivergence}[2]{ {\rm Div}_{#2} \left[ \, {#1} \, \right] }

\newcommand{\gradient}[1]{ {\rm \nabla} \left[ \, {#1} \, \right] }
\newcommand{\surfacegradient}[2]{ {\rm \nabla}_{#2} \left[ \, {#1} \, \right] }
\newcommand{\Gradient}[1]{ {\rm Grad} \left[ \, {#1} \, \right] }
\newcommand{\surfaceGradient}[2]{ {\rm Grad}_{#2} \left[ \, {#1} \, \right] }

\newcommand{\curl}[1]{ {\rm{curl}} \left[ \, {#1} \, \right] }
\newcommand{\Curl}[1]{ {\rm{Curl}} \left[ \, {#1} \, \right] }
\newcommand{\sym}[1]{ {\rm sym} \left[ \, {#1} \, \right] }

\newcommand{\writeoverrightleftarrows}{\operatornamewithlimits{\rightleftarrows}}

\newcommand{\identity}{\mathds{1}}

\newcommand{\uchempot}[1]{ \mu^u_{#1}}
\newcommand{\etachempot}[1]{\mu^\eta_{#1}}

\newcommand{\Temperature}{T}

\newcommand{\diffusivity}{\mbox{${\rm D} \mskip-8mu  | \,$}}
\newcommand{\mobility}{\mbox{${\rm u} \mskip-8mu  | \,$}}

\renewcommand{\em}[1]{\it{#1}}

\begin{document}

\title{A framework for modeling cells spreading, motility and the relocation of proteins on advecting lipid membranes } 

\author[1,3]{M. Serpelloni } \ead{m.serpelloni002@unibs.it}
\author[1,3]{M. Arricca} \ead{m.arricca@unibs.it}
\author[2,3]{C. Bonanno} \ead{c.bonanno@unibs.it}
\author[1,3]{A. Salvadori \corref{cor1}} \ead{alberto.salvadori@unibs.it}
\cortext[cor1]{Corresponding author}
\address[1]{Universit\`{a} degli Studi di Brescia, Department of Mechanical and Industrial Engineering, Brescia, 25123, Italy}
\address[2]{Universit\`{a} degli Studi di Brescia, Department of Civil, Environmental, Architectural Engineering and Mathematics, Brescia, 25123, Italy}
\address[3]{The Mechanobiology research center at Universit\`{a} degli Studi di Brescia , Brescia, 25123, Italy}
%

\begin{abstract}

The response of cells during spreading and motility is dictated by several multi-physics events, which are triggered by extracellular cues and occur at different time-scales. 
For this sake, it is not completely appropriate to provide a cell with classical notions of the mechanics of materials, as for ``rheology'' or ``mechanical response''. Rather, a cell is an alive system
with constituents that show a reproducible response, as for the {\em{contractility}} for single stress fibers or for the mechanical response of a biopolymer actin network, but that reorganize in response to external cues in a non-exactly-predictable and reproducible way.
Aware of such complexity, in this note we aim at formulating a multi-physics framework for modeling cells spreading and motility, accounting for the relocation of proteins on advecting lipid membranes.

\end{abstract}

\maketitle

\section{Introduction}
\label{sec:intro}

Receptors dynamic along cell membrane is a key factor in several biological phenomena, as for angiogenesis, tumor metastasis, endocytosis and exocytosis.
Angiogenesis is a multistep process in which endothelial cells are affected by several extracellular stimuli, including growth factors, extracellular matrix, and parenchymal and stromal cells. In this process, growth factor receptors as well as adhesion receptors convey the extracellular signaling in a coordinate intracellular pathway promoting cell proliferation, migration, and their reorganization in active vessels \cite{BentleyChakravartula}.  
Integrins are a family of cell adhesion receptors that support and modulate several cellular functions required for tumor metastasis. They can directly contribute to the control and progress of metastatic dissemination. During tumor development, changes in this family of receptors impact upon the ability of tumor cells to interact with their environment and enable metastatic cells to convert to a migratory and invasive phenotype. Integrins regulate each step of the metastasis and affect tumor cell survival and interaction with changing environments in transit from the primary tumor to distant target organs \cite{Felding-Habermann:2003aa}.
Receptor-mediated endocytosis is a process by which cells absorb metabolites, hormones, proteins – and, in some cases, viruses – by the inward budding of the plasma membrane (invagination). This process forms vesicles containing the absorbed substances and is strictly mediated by receptors on the surface of the cell \cite{StillwellBook2016Chapter17}.

Whereas uncountable papers have been published on the biology of cells spreading, motility and the relocation of proteins on advecting lipid membranes, the mathematical modeling definitely lags behind experiments and overall received much less attention. Although nowadays a widespread literature in mechanobiology exists, the relocation of proteins and their interaction with the reorganizing cytoskeleton in the biological phenomena mentioned above is still an ongoing research topic, let alone the formulation of efficient algorithms and computational solvers for three-dimensional simulations. 

In this note, we attempt at defining a multi-physics scheme for the modeling of cells spreading, motility and the relocation of proteins on advecting lipid membranes, framing the mathematical setting within the mechanics and thermodynamics of continua \cite{GurtinFriedAnand}, stemming from seminal works 
\cite{FreundLinJMPS2004, Shenoy2005, Deshpande2006} and accounting for recent literature, either connected to the endocytosis of virus in human and animal cells \cite{Gao2014, Gao2016, WiegoldEtAlPAMM2019} or ligand-receptor mediated raft formation \cite{CarotenutoJMPS2020}, chemotaxis \cite{BubbaEtAl2020}, surface-associated caveolae mechanotransduction \cite{LibermanEtAl2019}.

The paper is designed as follows. After a nomenclature of the main symbols and the definition of operators in a Lagrangian setting, 
the paper focuses in section \ref{sec:Relocation} upon the relocation and reaction of receptors on a lipid membrane that advects. The topic is purposely presented in a broad sense, in order to be applicable to several possible
receptors-ligands interactions: specific applications - carried out in \cite{DamioliEtAlSR2017}, \cite{SerpelloniEtAl2020} and in the companion paper \cite{salvadori_in_preparation} - deals with the relocation of vascular endothelial growth factor receptors and  integrins during endothelial cell adhesion and spreading. In spite of the generality, section \ref{sec:Relocation} is self-contained and includes the description of Reynold's theorem on a surface that advects, of the equations that rule proteins transport on an advecting lipid membrane, and eventually of the receptors-ligand interactions, in form of chemical reactions, that take place concurrently with relocation. A rather similar approach has been taken in section \ref{sec:ActinRelocation}, which concerns the relocation and reaction of actin to form biopolymers within the cytosol. The mechanical evolution of the cell is discussed afterwards in section \ref{sec:forcesandmomentum}: besides stating the classical balance laws (of linear and angular momentum), the section is accompanied by an extensive discussion on boundary conditions, aimed at showing that Neumann type of conditions, due to electrostatic interactions, are most likely not responsible for cell spreading and motion in view of the modest amount of energy involved in those interactions compared to the bulk energy of a cell. We concluded therefore that spreading is a result of extensional and contractile forces exerted by pseudopodia and the cytoskeleton machinery \cite{Reinhart-King2005}. Those forces have been investigated further in section \ref{sec:thermodynamics}, where the thermodynamics of receptors motion on the membrane was studied at first up to the constitutive theory and the receptors-ligand interactions kinetics. The analysis of the thermo-chemo-mechanics of cells is the last section of this work: in it, we highlight the role of strain and stress decompositions in order to model cell adhesion, protrusion, and contractility. A bibliographic review is presented in a rather extensive paragraph, showing various approaches pursued in the literature to cover the multiscale scenario of cell viscoelasticity and identifying missing pieces within the theoretical framework that we set in the present note.

\section{Nomenclature}
\label{sec:Nomenclature}

\subsection{Notation}

Vectors $\vect{a}$ will be denoted by an over-right-arrow, second order tensors $\tensor{A}, \tensor{a}$ by bold face. This notation does not apply to operators.

\subsection{Operators}
\noindent
- the symbol $\divergence{-}$ denotes the divergence operator in the current configuration, i.e.  $\divergence{ \vect{f} } = { \partial f_i} / { \partial x_i} $ \\
- the symbol $\Divergence{-}$ denotes the referential divergence operator, i.e.  $\Divergence{ \vect{f} } = { \partial f_i} / { \partial X_i} $ \\
- the symbol $\gradient{-}$ denotes the gradient operator in the current configuration \\
- the symbol $\Gradient{-}$ denotes the referential gradient operator  \\
- the symbol $\cdot$ denotes the single contraction of two vectors \\
- the symbol $:$ denotes the double contraction of two tensors \\
- the symbols $|| \vect{x} ||^2 $, $|| \tensor{x} ||^2 $ denote the squared norm of vector $\vect{x}$ or tensor  $\tensor{x}$ \\
- the symbol $^T$ denotes transposition of a tensor \\
- the symbol $^{-1}$ denotes the inverse of a tensor \\

\subsection{Variables and fields}
\noindent
- the symbol $t$ denotes time\\
- the symbol $\Omega(t) \in \mathbb{R}^3$ denotes  a volume that advects \\
- the symbol $\partial \Omega(t)$ denotes the surface of  $\Omega(t) $ \\
- the symbol ${\cal P}(t) \subset \partial \Omega(t)$ denotes a part of  $\partial \Omega(t)$ \\
- the symbol $\vect{v}_{adv}({\vect{x}}, t)$ denotes the velocity of advection at place ${\vect{x}}$ and time $t$ \\
- the symbol $\vect{n}({\vect{x}}, t)$ denotes the outward normal at place ${\vect{x}}$ and time $t$ \\
- the symbol $\tensor{l}({\vect{x}}, t)$ denotes the velocity gradient at place ${\vect{x}}$ and time $t$ \\
- the symbol $\tensor{d}({\vect{x}}, t)$ denotes the stretching at place ${\vect{x}}$ and time $t$ \\
- the symbol $\tensor{F}({\vect{X}}, t)$ denotes the deformation gradient at point ${\vect{X}}$ and time $t$ \\
- the symbol $\tensor{C}({\vect{X}}, t)$ denotes the right Cauchy-Green tensor at point ${\vect{X}}$ and time $t$ \\
- the symbol $\tensor{P}({\vect{X}}, t)$ denotes the first Piola stress tensor at point ${\vect{X}}$ and time $t$ \\
- the symbol $J({\vect{X}}, t)$ denotes the determinant ${\rm det}[ \tensor{F} ]$ at point ${\vect{X}}$ and time $t$ \\
- the symbol $\vect{n}_R({\vect{X}}, t)$ denotes the outward normal at point ${\vect{X}}$ and time $t$ \\
\noindent
- the symbol $m_a$ denotes the molar mass of species $s$ \\
- the symbol $c_a$ denotes the molarity of species $s$ \\
- the symbol $\rho_a$ denotes the density of species $s$ \\
- the symbol ${\overline s}_a$ denotes the mass supply of species $s$ \\
- the symbol ${s}_a$ denotes the molar supply of species $s$ \\
- the symbol $\vect{\hbar}_a$ denotes the density flux of species $s$ \\
- the symbol $\vect{h}_a$ denotes the molar flux of species $s$ \\

\begin{figure}[h]
\begin{subfigure} {0.5\textwidth}
  \includegraphics[height=8cm]{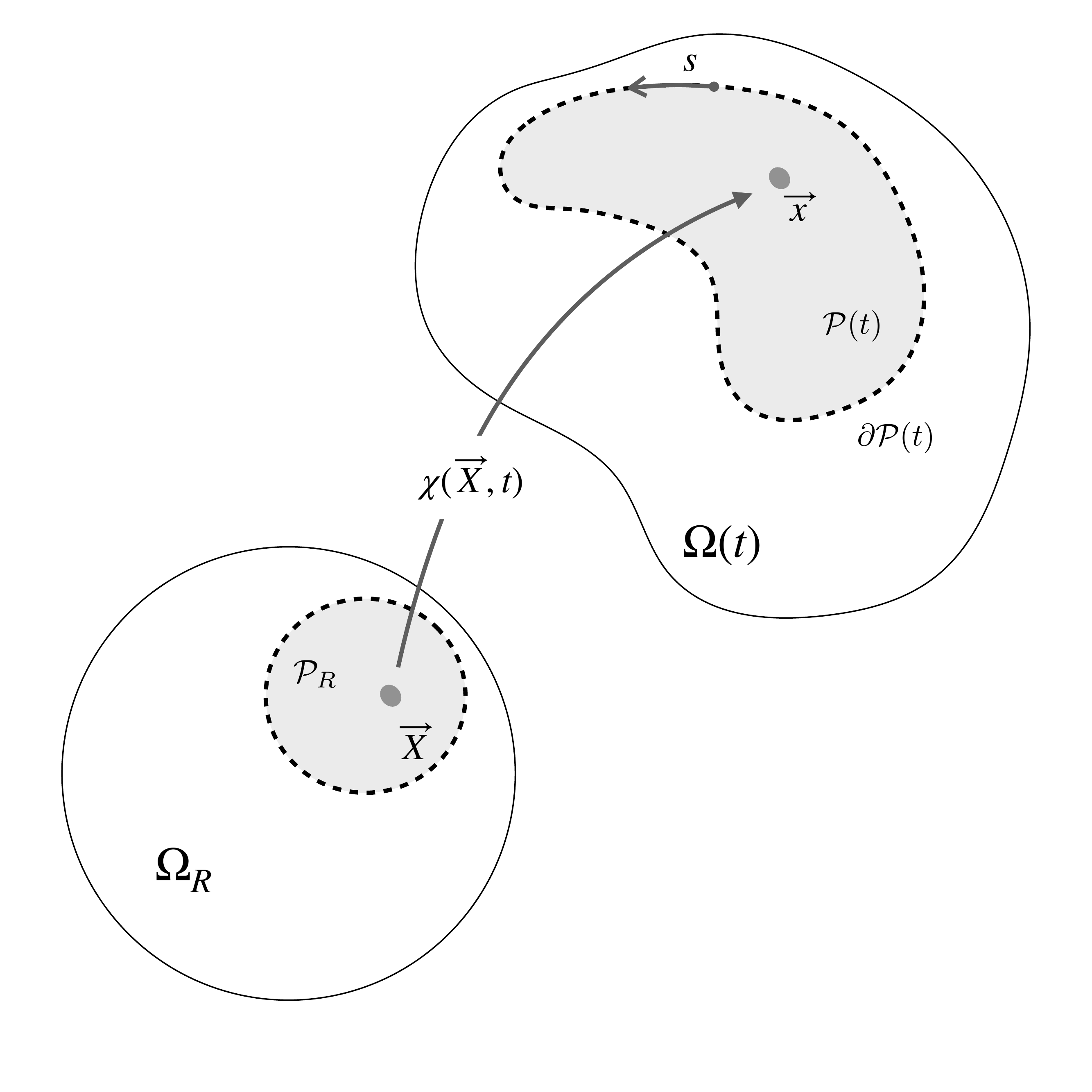}
\caption{The reference body $\Omega_R$ and the deformed body $\Omega(t)$. Note that  $\vect{x} \in {{\cal P}_(t)}  $ implies  $\vect{X} \in {{\cal P}_R}  $. }
\end{subfigure}
\begin{subfigure} {0.5\textwidth}
  \includegraphics[height=8cm]{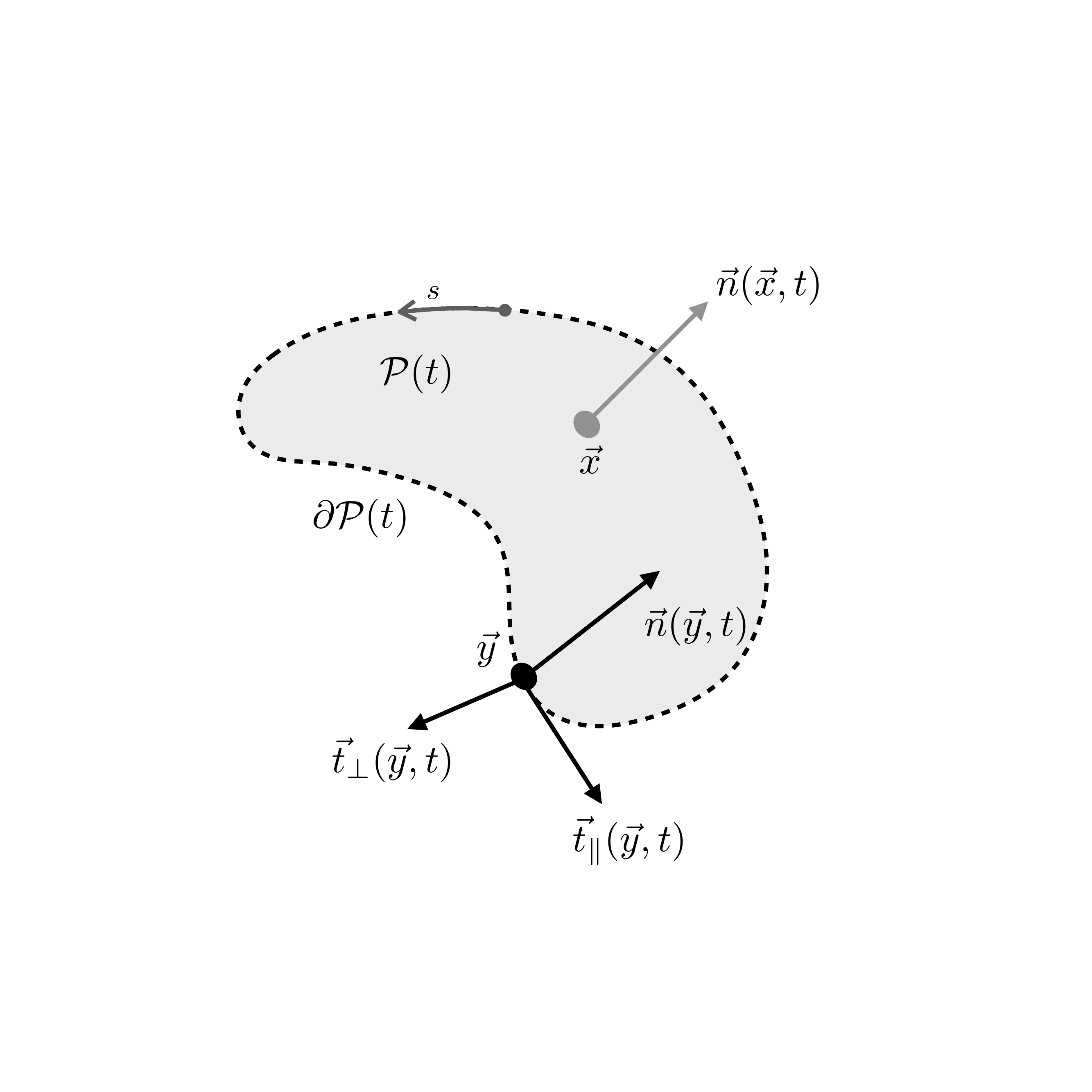}
\caption{ Frenet frame at point $\vect{y} \in \partial {\cal P}(t)$ and the normal vector $\vect{n}$ at point $\vect{x}  \in {\cal P}(t)$.}
\end{subfigure}
\caption{Notation}
\label{fig:Notation}
\end{figure}

\section{Definitions }
\label{sec:Definitions}

%

Denote with $\Omega(t)$ a volume that advects, and with $\partial \Omega(t)$ its surface. A point $\vect{x} \in \Omega(t)$ is defined as the image of a point $\vect{X}$ in a reference configuration $\Omega_R$
through a smooth function ${\chi}(\vect{X},t)$ termed {\em{motion}}. Following \cite{GurtinFriedAnand}, we will name {\em{deformation}} the snapshot of a motion at a fixed time $t$:
$$
{{\chi}}_t(\vect{X}) = {\chi}(\vect{X},t)
\; .
$$
The deformation is assumed to be a one-to-one map. In addition, denoting the deformation gradient with 
$$
 \tensor{F} =   {\Gradient{\chi_t}} 
 \; ,
$$
the requirement $J = \determinant{\tensor{F}} > 0$ holds.
Define on the surface a part ${\cal P}(t) \subset \partial \Omega(t)$ as in Fig. \ref{fig:Notation}, and consider a scalar function $f({\vect{x}}, t)$ with  ${\vect{x}} \in {\cal P}(t)$. 
Denote with 
$$\vect{v}_{adv}({\vect{x}}, t) = {\rm d}\vect{x} / {\rm d} t $$ 
the velocity of advection at location ${\vect{x}}$ and time $t$; such a velocity has an arbitrary direction, i.e. it is not necessarily tangent to $\partial \Omega(t)$.

The {\it Frenet-Serret} reference frame at a generic point $\vect{y} \in \partial {\cal P}(t)$ is defined as in Fig. \ref{fig:Notation}, in terms of the two unit vectors $\vec{t}_{\|}(\vec{y},t)$ (tangent) and $\vec{t}_{\bot}(\vec{y},t)$ (normal).
The vector $\vec{n}(\vec{y},t)$ (binormal) is here taken of non-unit length, being the imagine in $\Omega(t)$ of a unit vector  $\vec{n}_R$ in the reference configuration $\Omega_R$, by means of the contravariant transformation 
$$
 \vec{n} = \tensor{F}^{-T} \, \vec{n}_R 
 \; .
$$
On the other hand, the following covariant transformations hold:
$$
 \vec{t}_{\|  _R } = \tensor{F}^{-1}  \, \vec{t}_{\|} 
 \; ,
 \qquad
 \vec{t}_{\bot  _R} = \tensor{F}^{-1}  \, \vec{t}_{\bot} 
 \; ,
$$
with the obvious implication that $ \vec{t}_{\|  _R }$ and $ \vec{t}_{\bot  _R}$ are not unit vectors.
The Frenet formulae holds, namely:
$$
 \kappa \; \vec{t}_{\bot} = - \partial \vec{t}_{\|} / \partial s 
 \; ,
 \qquad
 \tau \; \vec{t}_{\bot} =  \partial \frac{\vec{n}}{|\vec{n}|} / \partial s 
 \; ,
 \qquad
 \kappa \; \vec{t}_{\|}  -  \tau \;   \frac{\vec{n}}{|\vec{n}|}  = \partial \vec{t}_{\bot} / \partial s  
 \; ,
$$
where $\kappa$ denotes the curvature and $\tau$ the torsion.

\bigskip
The {\it projected gradient operator} of a scalar field $f$ on a surface $\cal P$ is defined as follows 
\begin{subequations}
\begin{align}
\label{eq:graddef}
\surfacegradient{ f }{ {\cal P} }
=
\gradient{ f }
-
\; \frac{ \vect{n} \cdot \gradient{ f } }{ | \vect{n} |^2  }  \, \vect{n} 
\; ,
\end{align}
in the current configuration, whereas in the reference configuration it reads
\begin{align}
\label{eq:Graddef}
\surfaceGradient{ f }{ {\cal P} }
=
\Gradient{ f }
-
\;  \vect{n}_R \cdot \Gradient{ f }  \, \vect{n}_R 
\; ,
\end{align}
\end{subequations}
The {\it projected divergence operator} of a vector field $\vect{v}$, which has an arbitrary direction, on a surface $\cal P$ is defined as follows
\begin{subequations}
\begin{align}
\label{eq:divdef}
&
\surfacedivergence{  \vect{v}}{ {\cal P} }
=
\divergence{  \vect{v} }
-
\; \frac{ \vect{n} \cdot \tensor{l}  \vect{n} }{ | \vect{n} |^2  }
\; ,
\\
&
\label{eq:Divdef}
\surfaceDivergence{  \vect{v}_R}{ {\cal P}_R }
=
\Divergence{  \vect{v} }
-
\; { \vect{n}_R \cdot {\Gradient{\vect{v}_R}}  \vect{n}_R }
\; ,
\end{align}
\end{subequations}
in the current and reference configurations, respectively. Tensor $\tensor{l} $ is the gradient of $\vect{v}$,  $ \tensor{l}  = \gradient{ \vect{v} }$. Note that $\tensor{l}$ in eq. \eqref{eq:divdef} can be replaced by its symmetric part $\tensor{d} = \sym{ \tensor{l} } $,  since for any skew-symmetric tensor $\tensor{w} $ it holds $ \vect{n} \cdot \tensor{w}  \vect{n} = 0$ .
Alternative forms for the projected divergence operators are
\begin{subequations}
\begin{align}
\label{eq:divdef2}
\surfacedivergence{  \vect{v}}{ {\cal P} }
=
\curl{  \frac{ \vect{n}  }{ | \vect{n} |  } \times \vect{v} } \cdot   \frac{ \vect{n}  }{ | \vect{n} |  }
\; ,
\qquad
\surfaceDivergence{  \vect{v}}{ {\cal P}_R }
=
\Curl{  \frac{ \vect{n}_R  }{ | \vect{n}_R |  } \times \vect{v}_R } \cdot   \frac{ \vect{n}_R  }{ | \vect{n}_R |  }
\; .
\end{align}
\end{subequations}

\bigskip
\noindent
Provided sufficient smoothness, the divergence theorem holds also for advecting membranes, in the form: 
\begin{equation}
\label{eq:DivergenceTheorem}
\int_{{\cal P}(t)} \, 
\surfacedivergence{  \vect{g} }{ {\cal P} }
 \;  {\rm d} a
=
 \int_{\partial {\cal P}(t)} 
\,
\vect{g}   \cdot  \vect{t}_\bot
\,
{\rm d} \ell 
\; .
\end{equation}
The proof of this theorem, as well as for all other theorems not explicitly stated in this paper, can be found in \cite{MattiaThesis}.

\section{Relocation and reaction of receptors on a lipid membrane that advects }
\label{sec:Relocation}

\subsection{Reynold's theorem on a surface that advects }
\label{sec:Reynolds}

Reynold's theorem on ${\cal P}(t)$ reads as follows \cite{MattiaThesis}:
\begin{equation}
\label{eq:ReynoldsAdvectingSurface}
\frac{ {\rm d}}{ {\rm d} t} \int_{{\cal P}(t)} \, f  \, {\rm d} a 
=
\int_{{\cal P}(t)} 
\,
\frac{ \partial f}{ \partial  t}
\, 
+
 \; \surfacedivergence{ f \,  \vect{v}_{adv} }{\cal P}
\; 
{\rm d} a 
\; ,
\end{equation}
where $\vect{v}_{adv}({\vect{x}}, t)$ is the velocity of advection at location ${\vect{x}}$ and time $t$.  
By taking $f=1$, eq. \eqref{eq:ReynoldsAdvectingSurface}  depicts the area evolution of ${{\cal P}(t)}$as
$$
\frac{ {\rm d}}{ {\rm d} t} \int_{{\cal P}(t)}   \, {\rm d} a 
=
\int_{{\cal P}(t)} 
\,
 \; \surfacedivergence{   \vect{v}_{adv} }{\cal P}
\; 
{\rm d} a 
\; .
$$
It is intuitive that advection with velocity in the tangent plane has the potential of modifying the surface area, however 
even $\vect{v}_{adv} ({\vect{x}}, t)  \propto   \vect{n}({\vect{x}}, t) $ can do so, as for the homothetic expansion of a rubber balloon.
Reynold's theorem \eqref{eq:ReynoldsAdvectingSurface} can be also restated as
\begin{align}
\label{eq:ReynoldsAdvectingSurface1}
\frac{ {\rm d}}{ {\rm d} t} \int_{{\cal P}(t)} \, f({\vect{x}}, t)  \, {\rm d} a 
=
\int_{{\cal P}(t)} \, 
\frac{ {\rm d} \,  f({\vect{x}}, t) }{ {\rm d} t}   \,  
+
\, f({\vect{x}}, t) 
\;
\surfacedivergence{  \vect{v}_{adv}}{ {\cal P} }
 \;  {\rm d} a
\; .
\end{align}
and is a restriction on surfaces of the classical Reynold's transport relation on volumes ( see \cite{GurtinFriedAnand}, section 16 among others ).

\subsection{Mass transport on a surface that advects }
\label{sec:Balance}

\subsubsection{Mass balance in the current configuration for a convecting species }
\label{sec:MassCurrent}

Consider a generic species $a$ at a point $\vect{x}$ on the surface $\partial \Omega(t)$. Species $a$ convects with velocity $\vect{v}_a(\vect{x},t)$. The latter entails the dragging, or advection, velocity $\vect{v}_{adv}(\vect{x},t)$ and another velocity that is due to many possible physics, as for diffusion or migration. If {\em{internalization of species from the membrane}} is not allowed, the net velocity $ \vect{v}_a - \vect{v}_{adv}$ lays in the tangent plane of the membrane and 
\begin{equation}
\label{eq:tangentflux}
( \vect{v}_a - \vect{v}_{adv} ) \cdot \vect{n} = 0
\; .
\end{equation}
Since species are modeled on a membrane, which is a two-dimensional manifold, the surface density $\rho_a$ of species $a$ measures the mass of the species per unit surface. The {\it density} flux vector of species $a$, denoted with $\vect{\hbar}_a$, is the product of the surface density times the net velocity of species $a$, i.e.
\begin{equation}
\label{eq:fluxDef}
 \vect{\hbar}_a = \rho_a \; ( \vect{v}_a - \vect{v}_{adv} )
 \; .
\end{equation}
Define on the surface a part ${\cal P}(t) \subset \partial \Omega(t)$ as in Fig. \ref{fig:Notation}.
The flux of species $a$ across the boundary $\partial {\cal P}(t)$ is
$$
 \int_{ \partial {\cal P}(t) }  \, \vect{\hbar}_a \cdot \vect{t}_\bot \; {\rm d}\ell
$$
and the mass balance of species $a$ in the advecting configuration ${\cal P}(t)$ reads
\begin{equation}
\label{eq:rho:massbalancePt}
\frac{ {\rm d}}{ {\rm d} t} \int_{{\cal P}(t)} \, \rho_a({\vect{x}}, t)  \, {\rm d} a 
+
 \int_{ \partial {\cal P}(t) }  \, \vect{\hbar}_a \cdot \vect{t}_\bot \; {\rm d}\ell
=
\int_{{\cal P}(t)} \, {\overline s}_a({\vect{x}}, t)  \, {\rm d} a 
\; ,
\end{equation}
where ${\overline s}_a({\vect{x}}, t)$ is the surface mass supply\footnote{As an example, in biology cells may produce proteins that  move to the lipid membranes from the cytosol. } of species $a$. 
By means of the divergence theorem \eqref{eq:DivergenceTheorem} and of Reynold's transport theorem in the form \eqref{eq:ReynoldsAdvectingSurface1}, balance law \eqref{eq:rho:massbalancePt} becomes
\begin{equation*}
\label{eq:rho:massbalancePt2}
\int_{{\cal P}(t)} 
\,
\frac{ {\rm d} \rho_a }{ {\rm d}  t}
\, 
+
 \rho_a \; \surfacedivergence{  \vect{v}_{adv} }{ {\cal P} } 
+
\surfacedivergence{  \vect{\hbar}_a  }{ {\cal P} } 
\; 
{\rm d} a 
=
\int_{{\cal P}(t)} \, {\overline s}_a({\vect{x}}, t)  \, {\rm d} a 
\; .
\end{equation*}
Since it holds for all ${\cal P}(t)$, it eventually localizes as
\begin{equation}
\label{eq:rho:massbalancePtLocal}
\,
\frac{ {\rm d} \rho_a }{ {\rm d}  t}
\, 
+
 \rho_a \; \surfacedivergence{  \vect{v}_{adv} }{ {\cal P} } 
+
\surfacedivergence{  \vect{\hbar}_a  }{ {\cal P} } 
\; 
=
\, {\overline s}_a({\vect{x}}, t) 
\; .
\end{equation}
This formulation of the mass conservation law has been considered also in \cite{MikuckiZhouSIAM2017}.
The mass balance can be finally written in terms of surface molarity $c_a$ (in moles or molecules per unit surface), by division by the molar or molecular mass ($m_a$) of species $a$. By denoting with $c_a = \rho_a / m_a$, $s_a ={ \overline s}_a / m_a$, and $\vect{h}_a  = \vect{\hbar}_a  / m_a$
the local balance \eqref{eq:rho:massbalancePtLocal} becomes
\begin{equation}
\label{eq:c:massbalancePtLocal}
\,
\frac{ {\rm d} c_a }{ {\rm d}  t}
\, 
+
c_a \; \surfacedivergence{  \vect{v}_{adv} }{ {\cal P} } 
+
\surfacedivergence{  \vect{h}_a  }{ {\cal P} } 
\; 
=
\, {s}_a({\vect{x}}, t) 
\; .
\end{equation}

\subsubsection{Mass balance in the reference configuration for a convecting species }
\label{sec:MassReference}

The mass balance \eqref{eq:c:massbalancePtLocal} can be rephrased in the reference configuration at point $\vect{X}$ and time $t$. To this aim, define the reference molarity of species $a$
as
\begin{equation}
\label{eq:caR}
{c_{a_R}}( {\vect{X}}, t)
=
\, c_a ( \, {\vect{x}}({\vect{X}}, t \,), t \, )  
\;  j({\vect{X}}, t)
\; ,
\end{equation}
the reference flux vector ${ \vect{h}}_{a  _R}( {\vect{X}}, t)$ and the reference mass supply $s_{a  _R}( {\vect{X}}, t)$ as
\begin{equation}
\label{eq:haR}
{ \vect{h}}_{a  _R}
=
\,  j  \, \tensor{F}^{-1} \;  {\vect{h}_a}( \, {\vect{x}}({\vect{X}}, t \,), t \, )  
\; ,
\qquad
s_{a  _R}
=
\,  j  \, s_a( \, {\vect{x}}({\vect{X}}, t \,), t \, )  
\; ,
\end{equation}
respectively, 
where \cite{GurtinFriedAnand, paolucci2016}:
\begin{equation}
\label{eq:dadaR}
j
= 
J \; |  \tensor{F}^{-T}  \vect{n}_R | 
=
J \;  \sqrt{  \vect{n}_R \cdot   \tensor{C}^{-1}  \vect{n}_R } 
\; .
\end{equation}
The referential form of the mass balance \eqref{eq:c:massbalancePtLocal} can be derived from the mass balance in the form \eqref{eq:rho:massbalancePt}, and reads 
\begin{equation}
\label{eq:c:massbalancePLocal}
\frac{ \partial {c_{a_R}}   }{ \partial t}  
+
\,  
\surfaceDivergence{ { \vect{h}}_{a  _R} } {{\cal P}_R}   
=
s_{a  _R}  
\; .
\end{equation}
For the sake of brevity, the proof has been here omitted, interested readers may find it in \cite{MattiaThesis}.

\subsection{Relocation and reaction }
\label{sec:BalanceWith Chem}

%
The association and formation of a protein complex follow a two-steps mechanism; the formation of an 
encounter complex, in which previously free proteins show few specific interactions and assume many 
orientations, and the evolution of the encounter complex in the final complex. The encounter complex, 
which therefore represents the ensemble of orientations of proteins, is mostly dominated by electrostatic 
interactions. Under certain conditions it evolves in the final complex, when protein perfectly match each 
other, otherwise it dissociates and proteins return to be free \cite{Ubbink2009, Selzer2001}.

The two steps mechanism which describes the formation of a protein complex reads:
\begin{equation} 
\label{encounter_complex} 
	{\rm R}+ {\rm L}  \writeoverrightleftarrows_{k_{-1}}^{k_1}   {\rm C}^*
	\writeoverrightleftarrows_{k_{-2}}^{k_2}   {\rm C}
\end{equation}
where $\rm{R}$ and $\rm{L}$ are the receptors $ ({\rm R}) $ and ligands $ ({\rm L}) $ free proteins, $\rm {C}^*$ represents the encounter complex and 
$\rm{C}$ is the final complex.
In Equation \eqref{encounter_complex}, $k_1$ and $k_{-1}$ are the rate of formation and dissolution of the 
encounter complex, $\rm{C}^*$, whereas $k_2$ and $k_{-2}$ are the forward and reverse rate 
constants for formation of the final complex, $\rm {C}$, from $\rm{C}^*$. 

Assuming that the formation of  the encounter complex occurs whenever $\rm{R}$ and $\rm{L}$ 
are separated by an encounter distance smaller than $r$,  then 
%
%
$k_1 = 2 \pi  [D({\rm{R}})+D(\rm{L})]$, $k_{-1} = 2 [D({\rm{R}})+D({\rm{L}})] r^{-2}$. Here 
$D(\rm{R})$ and $D(\rm{L})$ are the translational diffusion constants for protein motion in the 
membrane and the equilibrium constant for the encounter step, $K_d = \pi r^2$, represents the 
area of a disk of radius $r$ \cite{Bell618}.
If the concentration of ${\rm C}^*$ is smaller than the concentration of free proteins or final complexes, 
it is a good approximation to set ${\rm d}{C}^*/{\rm d} t=0$, leading to the  
binding-unbinding interaction
\begin{equation} 
\label{eq:chem_react}
	{\rm R}+ {\rm L}  \writeoverrightleftarrows_{k_b}^{k_f}   {\rm C}
\end{equation}
most commonly used 
\cite{Bell618}.
%
%
A similar approach has been taken in \cite{DamioliEtAlSR2017,salvadoriHindawi2018} for the relocation of VEGFR-2 receptors and in  \cite{SerpelloniEtAl2020} for integrins.
Coefficients $k_f$ and $k_b$ are the kinetic constants of the forward and backward reactions, respectively. The rate of reaction \eqref{eq:chem_react}, denoted with $ w^{\eqref{eq:chem_react}} $and measured in $[ \frac{mol}{m^{2}s}]$, quantifies the net formation of (C) on the advecting membrane as the difference between the forward and backward reactions.
Equation %
\eqref{eq:c:massbalancePLocal} shall be extended to account for the reaction \eqref{eq:chem_react} and tailored to species $a = R, L, C$.

\bigskip
Receptors (either free or bound into the complex) are distributed along the membrane together with other lipid species and proteins. They are assumed to freely move laterally, effects due to steric hindrance are not accounted for. 
The amount of proteins per unit area that can be placed at a membrane location $\vect{x}$ is thus limited by the actual  size of the protein itself.
This evidence ushers the definition of a saturation limit for the species, 
${c_{a}^{max}} ({\vect x},t)$.

\bigskip
During their life, cells and their membranes undergo major {\em{macroscopic}} mechanical deformations. Studies on the red blood cell \cite{evans1973} suggest that the membrane deformation occur at constant area, but this evidence does not appear to be supported by experiments in endothelial cells during spreading \cite{Reinhart-King2005}. Individual protein and phospholipid can easily move laterally within the membrane, which results in a very low shear stiffness.
The {\em{fluid mosaic model}} \cite{SingerNicholson1972} captures this evidence, adding a questionable high resistance to areal expansion.  
Indeed the mechanisms that are in charge of areal expansion during cell spreading are complex and involve the micro-structural topology\footnote{Multiscale investigations, however, fall out of the scope of the present work.} of the membrane (as for flattening of invaginated membrane domains \cite{SensTurnerPhysRevE2006}, i.e. the role of the caveolae as membrane surface repository readily made available for fast geometrical evolution as during filopodia extension).
The structure of the lipid membranes, however, induce to suppose that
the saturation concentration ${c_{a}^{max}} ({\vect x},t)$,  i.e. the maximum number of moles or molecules per unit area for any species $a$, remains unchanged in time in the current configuration. 
This choice in turn entails that the number of moles or molecules per unit area in the reference configuration is not constant and evolves in time following eq. \eqref{eq:caR}, i.e.
\begin{equation}
\label{eq:caRmax}
{c_{a_R}^{max}}( {\vect{X}}, t)
=
\, c_a^{max}({\vect{x}}({\vect{X}}, t), t)  \,  j({\vect{X}}, t)
\; .
\end{equation}
Accordingly, the value of the non-dimensional ratio between the concentration of species $a$ and its amount ${c_{a}^{max}} $ at saturation,  
\begin{equation}
\label{eq:vartheta}
\vartheta_a =  {c_{a}}/{c_{a}^{max}}
\end{equation}
in the current configuration remains unchanged in the reference configuration
\begin{equation}
\label{eq:varthetaRmax}
\vartheta_{a_R}( {\vect{X}}, t ) = \vartheta_a ( {\vect{x}}, t ) 
\; .
\end{equation}

\bigskip
The kinetics of reaction $ \eqref{eq:chem_react} $ is modeled as for ideal systems via the law of mass action  \cite{deGrootBook} 
\begin{equation}
\label{eq:mass_action}
	w^{(\ref{eq:chem_react})}= k_f \,\frac{\vartheta_L}{(1- \vartheta_L)} \,\frac{\vartheta_R}{(1- \vartheta_R)} - \, k_b \, \frac{\vartheta_C}{(1- \vartheta_C)}
	\; .
\end{equation}
At chemical equilibrium, as $ w^{(\ref{eq:chem_react})}=0$, the concentrations obey the relation 
\begin{equation}
\label{eq:eq_const}
\frac{  k_f  }{  k_b } = 
\frac{\vartheta_C^{\rm eq}}{(1- \vartheta_C^{\rm eq})} \, \frac{(1- \vartheta^{\rm eq}_R)}{\vartheta^{\rm eq}_R} \, \frac{(1- \vartheta^{\rm eq}_L)}{\vartheta^{\rm eq}_L} 
= K_{\rm eq}^{(\ref{eq:chem_react})} 
\end{equation}
which defines the constant of equilibrium $K_{\rm eq}^{(\ref{eq:chem_react})} $ of reaction \eqref{eq:chem_react}. 

\bigskip
Far from the saturation limit,  $(1- \vartheta_a) \sim 1$ for all $a$. Accordingly, the mass action law \eqref{eq:mass_action} simplifies as
\begin{equation}
\label{eq:mass_action_dilute}
	w^{(\ref{eq:chem_react})}= \tilde{k}_f  \, c_L \, c_R - \, \tilde{k}_b \, c_C 
\end{equation}
once the new constants 
$$
 \tilde{k}_f  = k_f   ( {c_{L}^{max}} {c_{R}^{max}}  )^{-1}
 \; , 
 \qquad
 \tilde{k}_b  = k_b   ( {c_{C}^{max}}  )^{-1}
$$
are defined.

%

\bigskip
The diffusion of receptors and the viscous evolution of the cell during adhesion and migration appear to be much slower than the interaction kinetics, 
i.e. the time required to reach chemical equilibrium is orders of magnitude smaller than the time-scale of other processes. 
For this reason,  thermodynamic equilibrium may be invoked in place of a transient evolution, thus inferring the constraint $ w^{(\ref{eq:chem_react})}=0$ to the concentrations of species at all times. 
Far from saturation, equating \eqref{eq:mass_action_dilute} to zero implies that
\begin{equation}
 \label{cond:concentration1}
{c_C} = \frac{c_R \, c_L}{\alpha}
\; ,
\end{equation}
having denoted with $\alpha$ the following constant:
\begin{equation}
 \label{eq:alpha}
\alpha 
= \frac{ \tilde{k}_b }{ \tilde{k}_f } 
= \frac{c^{ max}_{R} \, c^{ max}_{ L}}{ c^{ max}_{C}}  \; \frac{1}{ K_{\rm eq}^{(\ref{eq:chem_react})}  }
\; .
\end{equation}
In view of identity \eqref{cond:concentration1}, the two concentrations $c_R$ and $c_L$ describe the problem in full, and the concentration of the complex can be deduced a posteriori. 

\bigskip
In vivo experiments show that the complex molecules usually have a much smaller mobility than receptors, perhaps induced by their size. 
For in vitro experiments \cite{DamioliEtAlSR2017,salvadoriHindawi2018,SerpelloniEtAl2020}, ligands %
are prevented to flow onto the substrate: 
given that complex molecules result from the interaction with immobile ligands, they are macroscopically steady as well.  
Since receptors move along the membrane, reaction \eqref{eq:chem_react} traps mobile receptors and vice-versa \cite{SalvadoriEtAlJMPS2018}. 
In this work, analogously to \cite{LiuJMPS2007}, ligands and complex are assumed to be motionless, i.e. 
\begin{equation}
 \label{eq:zero:flux}
\vect{h}_{L} = \vect{h}_{C} = \vect{0} 
\; .
\end{equation}

\bigskip
The reaction rate $w^{\eqref{eq:chem_react}}({\vect{x}}, t) $, being a mass supply, shall transform as ${s}_a({\vect{x}}, t) $ according to eq. \eqref{eq:haR}.  The invariance of $\vartheta_a $ with the configuration and the analysis of the mass action law \eqref{eq:mass_action} imply that the forward and backward ``constants'', which encompass the dimensionality of  $w^{\eqref{eq:chem_react}}({\vect{x}}, t) $, are not actually constants in the reference configuration. They rather change with time and with point $\vect{X}$ according to 
\begin{equation}
\label{eq:kfbR}
k_{f _R}({\vect{X}}, t)
=
\,  j({\vect{X}}, t) \, k_{f }
\; ,
\qquad
k_{b _R}({\vect{X}}, t)
=
\,  j({\vect{X}}, t) \, k_{b}
\; \,
\end{equation}
with $ j({\vect{X}}, t)$ as in \eqref{eq:dadaR}. The equilibrium constant in the reference configuration, being the ratio of $k_{f _R}$ and $k_{b _R}$ remains independent upon the configuration. 
Eventually, the mass action law \eqref{eq:mass_action} in the reference configuration writes
\begin{equation}
\label{eq:mass_action_ref}
	w^{(\ref{eq:chem_react})}_R= k_{f_R} \,\frac{\vartheta_L}{(1- \vartheta_L)} \,\frac{\vartheta_R}{(1- \vartheta_R)} - \, k_{b_R} \, \frac{\vartheta_C}{(1- \vartheta_C)}
	\; .
\end{equation}
In view of all considerations made so far, 
the local form \eqref{eq:c:massbalancePLocal} of the mass balance 
specify as follows ( omitting the dependency upon ${\vect{X}}$ and $t$ ):
\begin{subequations}
\begin{align}
&
\label{eq:mass_balance_ref_R}
\frac{ \partial {c_{R_R}}   }{ \partial t}  
+
\,  
\surfaceDivergence{ { \vect{h}}_{R_R} } {{\cal P}_R}   
+
\, w^{(\ref{eq:chem_react})}_R
\; 
=
s_{R_R}  
\;
,
\\
&
\label{eq:mass_balance_ref_L}
\frac{ \partial {c_{L_R}}   }{ \partial t}  
+
\, w^{(\ref{eq:chem_react})}_R
\; 
=
0 
\;
,
\\
&
\label{eq:mass_balance_ref_C}
\frac{ \partial {c_{C_R}}   }{ \partial t}  
-
 \, w^{(\ref{eq:chem_react})}_R
\; 
=
0
\;
.
\end{align}
\label{eq:ref:threegoveq}
\end{subequations}
Equation \eqref{eq:mass_balance_ref_R} is defined on the membrane surface $\partial \Omega_R$, where the receptors flow. The supply $s_{R_R} $ accounts for internalization or generation of proteins: it  is the amount of receptors that are generated within the cell and reach the membrane or that internalize.  It can be related to the change in the membrane area through a parameter $\kappa_{R_R}$ as 
\begin{align}
\nonumber
s_{R_R}({\vect{X}}, t)
&
=
\kappa_{R_R}
\frac{\partial j}{\partial t}
\\
&
=
\kappa_{R_R}
\left[
 \; 
 | \tensor{F}^{-T} \, \vect{n}_R | \, J \,  \trace{\tensor{ l }}
 \,
 -
 \frac{J}{2} \, \frac{1}{ | \tensor{F}^{-T} \, \vect{n}_R | }
 \;
 \vect{n}_R \cdot \tensor{C}^{-1}  \, \frac{\partial \tensor{C} }{ \partial t} \, \tensor{C}^{-1} \,  \vect{n}_R 
\right]
\; .
\label{eq:sR:howitworks}
\end{align}
At all points at which ligands and receptors do not interact, the reaction rate $w^{(\ref{eq:chem_react})}_R$ vanishes. Equation \eqref{eq:mass_balance_ref_L} is rather defined in the location where ligands stand. In vitro, a given amount of ligands (which can be thought of as the initial condition of eq. \eqref{eq:mass_balance_ref_L}  are spread upon a microscope slide. Finally, eq.  \eqref{eq:mass_balance_ref_C} is defined in the contact zone between the cell and the slide where reaction (\ref{eq:chem_react}) takes place.

\bigskip
It is convenient to rephrase eq. \eqref{eq:mass_balance_ref_L}  in terms of the ``ligands made available for the reaction'' in place of the ``ligands spread on the slide".  
The former ligands are the ones ``felt''  at a point on the membrane as the distance from such a point and the substrate, where ligands are spread out, becomes sufficiently small.

Such a distance can be understood as a cutoff, within which the formation of an encounter complex, 
$\rm {C}^*$, becomes possible as a consequence of diffusion, as made clear in 
\cite{Ubbink2009, Selzer2001,Bell618, BongrandRPP1999}.
Despite the size of the cutoff distance remains inaccurately estimated, it was established to be on the order
of tens nanometers \cite{Bell618, FreundLinJMPS2004}. It arises form the interplay of attractive
and repulsive forces between either two cells or a cell and a substrate. Indeed, negative electrical charge carried by 
cells generates repulsive electrostatic forces - {\em repulsive barrier} - which is further enriched by an 
additional resistance provided by the compression of the glycocalyx proteins. Rather, electrodynamic
van der Waals forces are expected to be attractive \cite{Bell618}. Both van der Waals and compressive 
forces are characterized as non-specific long ranged forces, whereas cell adhesion is generally 
mediated by the specific short ranged receptor-ligand interactions, which can cause cell adhesion 
much more tightly than the non-specific electrical forces \cite{Bell618, LiuJMPS2007}. 
Cells separated by a distance less than, or equal to, the cutoff distance should form a zone of 
adhesion with the substrate by means of local fluctuations in receptors density, so that small regions 
of increased density can penetrate through the resisting potential to react with the source of ligands 
on substrate \cite{FreundLinJMPS2004}.

This point of view, which corresponds to the picture of tight receptor-ligand bond as a set of weak non covalent physical interactions \cite{Alberts2002}, is made explicit by a supply function $s_{L  _R} $, that vanishes at long ranges and rapidly reaches the initial concentration of ligands available for the reaction at short distances
\begin{align}
\label{eq:mass_balance_ref_L_final}
\frac{ \partial {c_{L_R}}   }{ \partial t}  
+
\, w^{(\ref{eq:chem_react})}_R
\; 
=
s_{L  _R}  
\;
.
\end{align}
The ligand supply $s_{L_R}({\vect{X}}, t)$ becomes available for the reaction during the spreading of the cell. It seems to be logically related to: i) a gap function between the substrate rich in ligands and the cell membrane {\it{in the current configuration}}; ii) a lag in time, namely a point-wise function of an internal variable that activates when the gap function is below some threshold and is related to the chemical kinetics of the binding-unbinding reaction \eqref{eq:chem_react}. 
In this form, all three equations \eqref{eq:mass_balance_ref_R}, \eqref{eq:mass_balance_ref_C}, \eqref{eq:mass_balance_ref_L_final} can be written on the membrane $\vect{X} \in \partial \Omega_R$.

\bigskip
Assuming that the time scale of the chemical reaction is much faster than other processes, the concentrations of species may be governed by thermodynamic equilibrium at all times. The concentration of complex $c_{C_R}$ relates then to the others by the equation $ w^{(\ref{eq:chem_react})}=0$, which leads to eq. \eqref{cond:concentration1} in the current configuration. Making use of mapping \eqref{eq:caR}, eq. \eqref{cond:concentration1} relates the concentration of complex in the reference configuration $c_{C_R}$ to the concentration of ligands and receptors in the same configuration $c_{L_R}$, $c_{R_R}$ as follows

\begin{subequations}

\begin{equation}
 \label{cond:ref:concentration1}
{c_{C_R}} = \frac{c_{R_R} \, c_{L_R}}{\alpha_R({\vect{X}}, t)}
\; ,
\qquad
\alpha_R ({\vect{X}}, t) = \alpha  \,  j({\vect{X}}, t)
\; ,
\end{equation}
with constant $\alpha$ defined in eq. \eqref{eq:alpha}. Transformation \eqref{cond:ref:concentration1} is consistent with the assumption \eqref{eq:caRmax} made on how saturations transform.

\bigskip
In conclusion, exploiting identity \eqref{cond:ref:concentration1}, the two concentrations $c_{R_R}$ and $c_{L_R}$ fully describe the problem in the assumption of infinitely fast kinetics, whereas the concentration of the complex can be deduced a posteriori. The two governing equations descend from eqs.\eqref{eq:ref:threegoveq} and read:
\begin{align}
&
\frac{ \partial {c_{R_R}}   }{ \partial t}  
+
\, 
\frac{ \partial {c_{C_R}}   }{ \partial t}  
+
\,  
\surfaceDivergence{ { \vect{h}}_{R_R} } {{\cal P}_R}   
\; 
=
s_{R_R}  
\;
,
\qquad
\vect{X} \in \partial \Omega_R 
\;
,
\\
&
\frac{ \partial {c_{L_R}}   }{ \partial t}  
+
\, 
\frac{ \partial {c_{C_R}}   }{ \partial t}  
\; 
=
s_{L  _R}  
\;
,
\qquad
\vect{X} \in \partial \Omega_R 
\;
.
\end{align}
\label{eq:ref:twogoveq}
\end{subequations}
Equations \eqref{eq:ref:twogoveq}, with associated initial conditions
\begin{eqnarray*}
 c_{R_R} ( \vect{X}, 0 ) = c^0_{R_R}( \vect{X} )  \; , \qquad
 c_{L_R} ( \vect{X}, 0 )  = 0  \; , \qquad
 c_{C_R} ( \vect{X}, 0 )  = 0   \; 
\end{eqnarray*}
and Dirichlet-Neumann boundary conditions 
define the relocation of receptors that undergo binding-unbinding reactions on the reference configuration of a membrane that advects.  
These are balance equations and as such hold for any constitutive behavior for the mass flux.
These equations are {\em{coupled to the mechanical evolution of the cell}}  (i.e. adhesion, spreading, migration) through the function $s_{L  _R} (\vect{X}, t)$,
which ``transfers'' ligands on the membrane according to the geometry of the cell.

\section{Relocation and reaction of actin to form biopolymers }
\label{sec:ActinRelocation}

The extensive mathematical description made in section \ref{sec:Relocation} will guide the modeling of the relocation and reaction of actin to form biopolymers
in the cytosol, which will be summarized here in a shorter shape.

Biopolymers are composed of actin, a protein termed globular or G-actin in its monomeric form and F-actin when it forms filamentous polymers. In turn, actin filaments can bundle to form stress fibers, or cross-link to form polymer networks that allow the movement of the cell. Polymerization is usually triggered by extracellular signals. In the case of cell locomotion, for instance, the cell extends finger-like protrusions by which the cell ``feels'' the surrounding surface. As done in \cite{deshpandeEtAlPRAS2007}, the precise details of the signaling pathways are here ignored. Rather, the level of signaling 
is assumed given in the reference configuration by a function 
\begin{equation} 
\label{eq:actin_signaling} 
  {\cal C} ( \vect{X}, t ) = \gamma_i  \,  \exp \left[{ - |  \vect{x}(  \vect{X}, t  ) -  \vect{y}_i | }  \right] \,  \exp \left[ { - \frac{ t-\tau_i}{ \theta } } \right]
\end{equation}
that accounts for the location of discrete signaling points $\vect{y}_i$ in the surroundings emitting signals of intensity $\gamma_i$  at time $\tau_i$; $\theta$ is the decay constant of the signal. This approach in modeling the external stimulus is similar to the membrane activator in \cite{Moure:2018aa}.

The transduction of the signal results in the polymerization of the actin
filaments and their cross-linking or bundling. The formation of single actin filaments can be modeled as a bimolecular reaction similar to \eqref{encounter_complex}, as in \cite{IntroductiontoCellMechanicsandMechanobiology}; in this note, the biopolymer turn-over will be described at a larger scale, involving the interplay between fundamental units and stress-fibers or pseudopodia, in the form
\begin{equation} 
\label{eq:actin_polymerization} 
	{\rm G} 
	\writeoverrightleftarrows_{k_b}^{k_f}   {\rm F}
\end{equation}
with ${\rm F}$ denoting either one of the two biopolymers. The network or fiber formation rate of reaction \eqref{eq:actin_polymerization}, denoted with $w^{(\ref{eq:actin_polymerization})}$, is influenced by mechanical stresses: stress fibers stability is favored by tension, for instance. For this reason, 
the stress tensor enters the chemical potential and the dissociation reaction of biopolymers. 
The kinetics of reaction $ \eqref{eq:actin_polymerization} $ is modeled via the law of mass action, properly extended to account for signaling:
\begin{equation}
\label{eq:actin_mass_action}
	w^{(\ref{eq:actin_polymerization})}  ( \vect{X}, t )=  {\cal C} ( \vect{X}, t ) \; k_f \,\frac{\vartheta_G}{(1- \vartheta_G)}  - \, {\cal{D}} (  \vect{X}, t ) \, k_b \, \frac{\vartheta_F}{(1- \vartheta_F)}
	\; ,
\end{equation}
having already discussed the meaning of the ratio $\vartheta$ in eq. \eqref{eq:vartheta}.  Function $\cal{D}$ accounts for the role of the stress in the dissociation of biopolymers, see for instance \cite{deshpandeEtAlPRAS2007}.

\subsection{Mass transport in the cytosol}
\label{sec:CytosolBalance}

%

Consider a generic species $a$ at a point $\vect{x}$ in the cytosol $\Omega(t)$. 
The mass balance of species $a$ in the advecting configuration ${\cal Q}(t)$ localizes as
\begin{equation}
\label{eq:rho:massbalanceQtLocal}
\,
\frac{ {\rm d} \rho_a }{ {\rm d}  t}
\, 
+
 \rho_a \; \divergence{  \vect{v}_{adv} }
+
\divergence{  \vect{\hbar}_a  }
\; 
=
\, {\overline s}_a({\vect{x}}, t) 
\; ,
\end{equation}
with  $ \vect{\hbar}_a$ and $ \vect{v}_{adv} $ defined earlier in section \ref{sec:MassCurrent}, $\rho_a$ is the density of species $a$. 
The mass balance can be restated in terms of molarity $c_a$ (in moles or molecules per unit volume), by division by the molar or molecular mass ($m_a$) of species $a$. By denoting with $c_a = \rho_a / m_a$, $s_a ={ \overline s}_a / m_a$, and $\vect{h}_a  = \vect{\hbar}_a  / m_a$
the local balance \eqref{eq:rho:massbalanceQtLocal} becomes
\begin{equation}
\label{eq:c:massbalanceQtLocal}
\,
\frac{ {\rm d} c_a }{ {\rm d}  t}
\, 
+
c_a \; \divergence{  \vect{v}_{adv} }
+
\divergence{  \vect{h}_a  }
\; 
=
\, {s}_a({\vect{x}}, t) 
\; .
\end{equation}
The latter can be rephrased in the reference configuration at point $\vect{X}$ and time $t$. To this aim, define the reference molarity of species $a$
as
\begin{equation}
\label{eq:JcaR}
{c_{a_R}}( {\vect{X}}, t)
=
\, c_a ( \, {\vect{x}}({\vect{X}}, t \,), t \, )  
\;  J({\vect{X}}, t)
\; ,
\end{equation}
the reference flux vector ${ \vect{h}}_{a  _R}( {\vect{X}}, t)$ and the reference mass supply $s_{a  _R}( {\vect{X}}, t)$ as \cite{GurtinFriedAnand}
\begin{equation}
\label{eq:JhaR}
{ \vect{h}}_{a  _R}
=
\,  J  \, \tensor{F}^{-1} \;  {\vect{h}_a}( \, {\vect{x}}({\vect{X}}, t \,), t \, )  
\; ,
\qquad
s_{a  _R}
=
\,  J  \, s_a( \, {\vect{x}}({\vect{X}}, t \,), t \, )  
\; ,
\end{equation}
respectively. The reaction rate $w^{\eqref{eq:actin_polymerization}}({\vect{x}}, t) $, being a mass supply, shall transform according to eq. \eqref{eq:JhaR}b.  The invariance of $\vartheta_a $ with the configuration and the analysis of the mass action law \eqref{eq:actin_mass_action} imply that the forward and backward ``constants'', which encompass the dimensionality of  $w^{\eqref{eq:actin_polymerization}}({\vect{x}}, t) $, are not actually constants in the reference configuration. They rather change with time and with point $\vect{X}$ according to 
\begin{equation}
\label{eq:kfbR}
k_{f _R}({\vect{X}}, t)
=
\,  J({\vect{X}}, t) \, k_{f }
\; ,
\qquad
k_{b _R}({\vect{X}}, t)
=
\,  J({\vect{X}}, t) \, k_{b}
\; \,
\end{equation}
The ratio $k_{f _R}/k_{b _R}$ remains independent upon the configuration. 
The referential form of the mass balance equations eventually reads 
\begin{subequations}
\begin{align}
&
\label{eq:c:mass_balance_ref_G}
\frac{ \partial {c_{G_R}}   }{ \partial t}  
+
\,  
\Divergence{ { \vect{h}}_{G  _R} }  
+
\, w^{(\ref{eq:actin_polymerization})}_R
\; 
=
s_{G _R}  
\;
,
\\
&
\label{eq:c:mass_balance_ref_F}
\frac{ \partial {c_{F_R}}   }{ \partial t}  
+
\,  
\Divergence{ { \vect{h}}_{F  _R} }  
-
\, w^{(\ref{eq:actin_polymerization})}_R
\; 
=
s_{F_R}  
\;
.
\end{align}
\label{eq:ref:actin_mass_balance_ref}
\end{subequations}
As for the complex molecules, filaments usually have a much smaller mobility than monomers and might be assumed to be motionless, i.e. 
\begin{equation}
 \label{eq:zero:filament_flux}
\vect{h}_{F} =  { \vect{h}}_{F  _R} = \vect{0} 
\; .
\end{equation}
The diffusion of monomers appears to be much slower than the interaction kinetics and
the concentrations of species may be governed by thermodynamic equilibrium at all times \cite{VernereyFarsad2014}. The concentration of filaments $c_{F_R}$ relates then to the monomers by the equation $ w^{(\ref{eq:actin_polymerization})}=0$, mediated by the local amount of signaling and stress. Equations \eqref{eq:ref:actin_mass_balance_ref}, with associated initial conditions
\begin{eqnarray*}
 c_{G_R} ( \vect{X}, 0 ) = c^0_{G_R}( \vect{X} )  \; , \qquad
 c_{F_R} ( \vect{X}, 0 )  = c^0_{F_R}( \vect{X} )   \; 
\end{eqnarray*}
and Dirichlet-Neumann boundary conditions 
define the relocation of monomers that undergo polymerization reactions in the reference configuration.

\section{Mechanical evolution of the cell }
\label{sec:forcesandmomentum}

Based upon the selection of the mechanisms that are supposed to govern the structural response of the cell, the balance laws of linear and angular momentum come out. Literature provides two basic approaches, whether the structural functions are demanded entirely to the cell membrane \cite{Joanny2013,Kruse2005,Prost2015,LaTorreEtAlNature2018,RahimiPRE2012} or to the development of a cytoskeletal structure within the bulk of the cell \cite{,deshpandeEtAlPRAS2007,Deshpande2006,DeshpandeEtAlJMPS2008,McEvoyJMBB2017,McMeekingDeshpande2017,PathakJAM2011,RonanEtAlBMM2014,RonanJMBB2012,VigliottiBMM2016}. The influence of curvature on the elastic stiffness of the membrane appears to be related to the size of the cell \cite{GolestanehBMM2016} and seems to be negligible for endothelial cells of diameter $\sim 10 \mu {\rm m}$.  These two evidences lead to consider the reorganization of the cytoskeleton through a network of actin and intermediate filaments and microtubules the main responsible for the mechanical response of endothelial cells, coupled to a passive behavior dictated by the viscosity of the cytosol as in \cite{deshpandeEtAlPRAS2007,Deshpande2006,VigliottiBMM2016}. Accordingly, balance of linear and angular momentum will be formulated for the bulk of the cell rather than the membrane.

\bigskip
Forces in continuum mechanobiology are described spatially by {\em{contact forces}} between adjacent spatial regions 
(as for the forces exchanged by the substrate and the cell during adhesion), {\em{surface forces}} exerted on the boundary of the cell by the environment 
(as for the receptor-ligand attractive interaction \cite{Bell618,Bell1984} and repulsive electrostatic interactions), {\em{body forces}} exerted on the interior points by the environment
(as for the gravity or pseudopodia forces that preside migration). Contact and surface forces, acting on $\partial \Omega(t)$ will be denoted henceforth with ${\vect t}(\vect{x},t )$ whereas body forces will be denoted with  ${\vect b}(\vect{x},t )$.
Their referential counterparts will inherit the subscript $_R$.

Throughout the rest of the paper we will neglect inertia forces, although some authors \cite{Allena:2013aa} pinpointed the role of inertia forces during migration. 
Accordingly, the balance of linear and angular momentum, which are assumed to hold at each time for all spatial regions ${\cal Q}(t) \subseteq \Omega(t)$,  read:
\begin{subequations}
\begin{align}
\label{eq:linearmomentum}
&
\int_{\partial {\cal Q}(t)}  {\vect t}(\vect{x},t ) \; {\rm d} a + \int_ {{\cal Q}(t)}  {\vect b}(\vect{x},t ) \; {\rm d} v  = \vect{0} 
\; ,
\\
\label{eq:angularmomentum}
&
\int_{\partial {\cal Q}(t)} {\vect{r}} \times {\vect t}(\vect{x},t ) \; {\rm d} a + \int_ {{\cal Q}(t)}  {\vect{r}} \times {\vect b}(\vect{x},t ) \; {\rm d} v  = \vect{0} 
\end{align}
\label{eq:lin_ang_momentum}
\end{subequations}
with $\vect{r}$ denoting the position vector with respect to an arbitrary pole. 
Classical arguments of continuum mechanics lead to localize eqs. \eqref{eq:lin_ang_momentum} in the reference configuration, in terms of the (first) Piola stress tensor $\tensor{P}$
and of the body forces measured per unit volume in the reference body 
$$  {\vect b}_R(\vect{X},t ) =  J(\vect{X},t ) \; {\vect b}(\vect{x}(\vect{X},t ),t )  \; .$$
The referential local form of the balance of linear momentum reads
\begin{subequations}
\begin{equation}
\label{eq:ref_linearmomentum}
\Divergence{ \tensor{P}} + \vect{b}_R = \vect{0}
\;
,
\qquad
\vect{X} \in  \Omega_R 
\; .
\end{equation}
The first Piola stress tensor $\tensor{P}$ must satisfy the local angular momentum balance 
\begin{equation}
\label{eq:ref_angularmomentum}
\tensor{P} \tensor{F}^T =  \tensor{F} \tensor{P}^T \; . 
\end{equation} 
\label{eq:ref_momentum}
\end{subequations}

\subsection{Boundary conditions}

Contact and surface forces are boundary conditions for problem \eqref{eq:ref_linearmomentum}. They emanate from electrostatic long or short range interactions, from receptor-ligand adhesion forces, as well as from contact tractions after adhesion.
A vast literature \cite{IntermolecularandSurfaceForces,CellBiologyByTheNumbers,IntroductiontoCellMechanicsandMechanobiology} has been devoted to quantify the forces involved in these interaction mechanisms. It emerges that uncertainties remain in the establishment of realistic values for attraction forces, not surprisingly due to the complexity of the required experimental tasks.

\bigskip
Studies on the influence of non-specific {\em{traction forces in cell adhesion}} were performed at different time scales, 
from minutes - as for the spreading of a mouse embryonic fibroblasts on a matrix-coated surface 
\cite{DubinThaler2004} - to several hours - as for a bovine aortic endothelial cells on polyacrylamide 
gels \cite{Reinhart-King2005} - for different cell sizes. Analyses refer mostly to the early 
stage of adhesion: as pointed out in \cite{LiuJMPS2007}, traction models are 
helpful  under specific conditions and particularly in predicting isotropic early stage of cell adhesion, which is essentially 
independent on cytoskeleton 
remodeling. Isotropic spreading is made possible by higher ligands 
densities; at lower densities of ligands, cells tend to spread anisotropically, by extending pseudopodia 
randomly along the cell membrane \cite{Reinhart-King2005}. This has been made clear also 
in modeling micropipette-manipulated red blood cell attachment-detachment from a substrate 
\cite{ChengJMPS2009}, which was performed in $\approx 50 \ ms$ showing that after approximately 
a third of the adhesion-spreading time, the adhesion-traction forces level off and to further 
increase spreading area, receptor diffusion from remote area of the cell to the spreading front is 
required. 

Roughly the same concept has been explored in \cite{SohailIJSS2013}, dealing with charged 
flexible particles that adhere to an oppositely charged rigid substrate 
due to electrostatic attraction forces. Surface forces drive the adhesion of small 
particles. The cell radius in the reference, unstressed configuration was considered in the micron/sub-micron range 
$1 \ \mu m$ in \cite{SohailIJSS2013} or even smaller $12.5 \ nm$ in \cite{GolestanehBMM2016}.  

According to \cite{Shenoy2005}, adhesion and spreading also require transport of receptors from the apical to the basal part of the cell in order to generate attractive forces.

\bigskip
In this paper we do not account explicitly for integrins, as done in \cite{RonanEtAlBMM2014} among others, yet we will use the approaches in \cite{RonanEtAlBMM2014, GolestanehBMM2016} to discuss the magnitude of {\em{traction forces in cell spreading}}. According to \cite{GolestanehBMM2016}, Neumann tractions emanate from short-range, noncovalent interactions between one receptor and one ligand due to polarization of a non-polar ligand molecule in the electrostatic field of a charged receptor. The binding force on the membrane per unit area in the current configuration was given as  
\begin{equation}
\label{eq:attractiveforces}
\vect{t}(\vect{x}) = - C ( K g_N + 1) \,  ( ( K g_N + 1)^2 + 1 )  \, g_N^{-5} \, \exp(-2 K g_N ) \, \rho_{rl}(\vect{x})  \ey
\end{equation}
where: $g_N$ is the gap between receptors and ligands, $\rho_{rl}(\vect{x})$ is the minimum concentration of receptors and ligands at location $\vect{x}$, $C$ is the number of weak noncovalent sub-bonds which form the interaction between one receptor and one ligand, $K$ is the inverse of the Debye length. It is of course particularly complex to provide parameters with high accuracy: assuming that the values provided in \cite{GolestanehBMM2016} apply also to endothelial cells, one would set $C = 1.17 \times 10^{-7}  {\rm fN } \mu{\rm m}^{-5}$, $K=1$. 

\begin{figure}[h]
\begin{subfigure} {0.5\textwidth}
  \includegraphics[width=8.5cm]{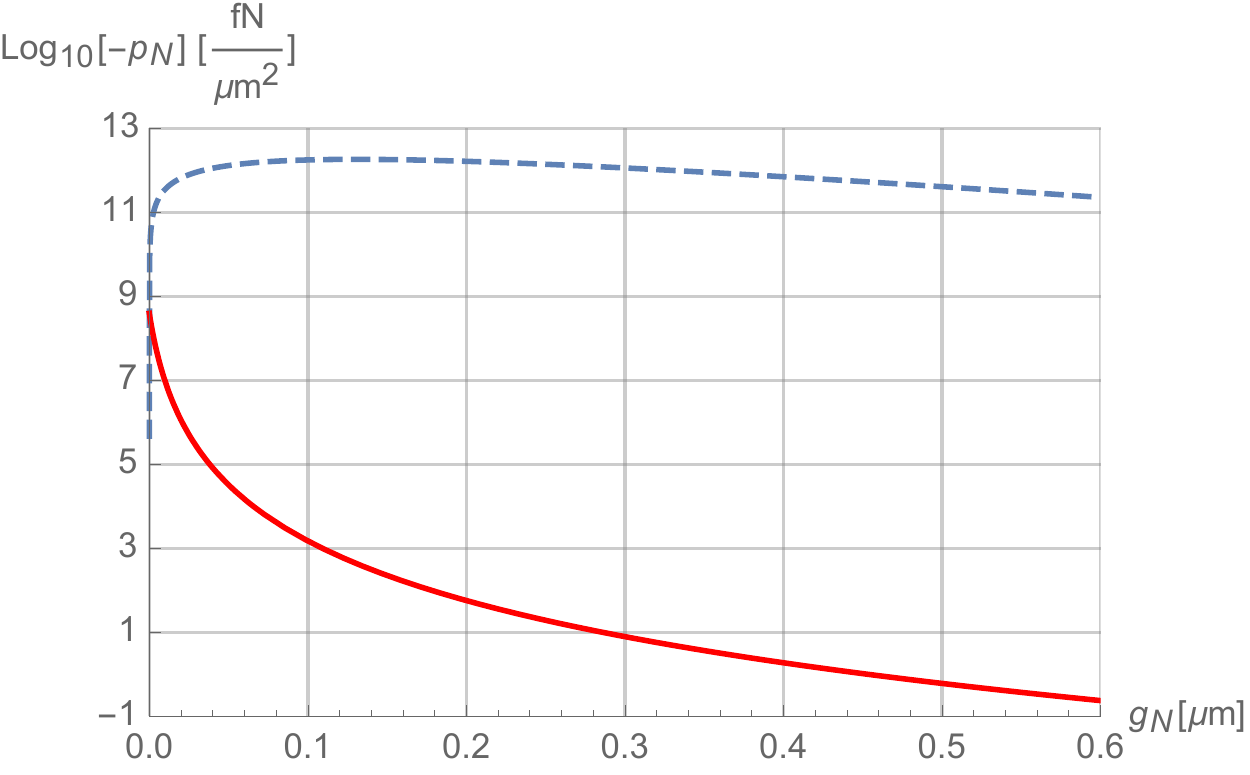}
\caption{According to  \cite{GolestanehBMM2016} (continuum) compared to  \cite{RonanEtAlBMM2014} (dashed) }
\end{subfigure}
\begin{subfigure} {0.5\textwidth}
  \includegraphics[width=8.5cm]{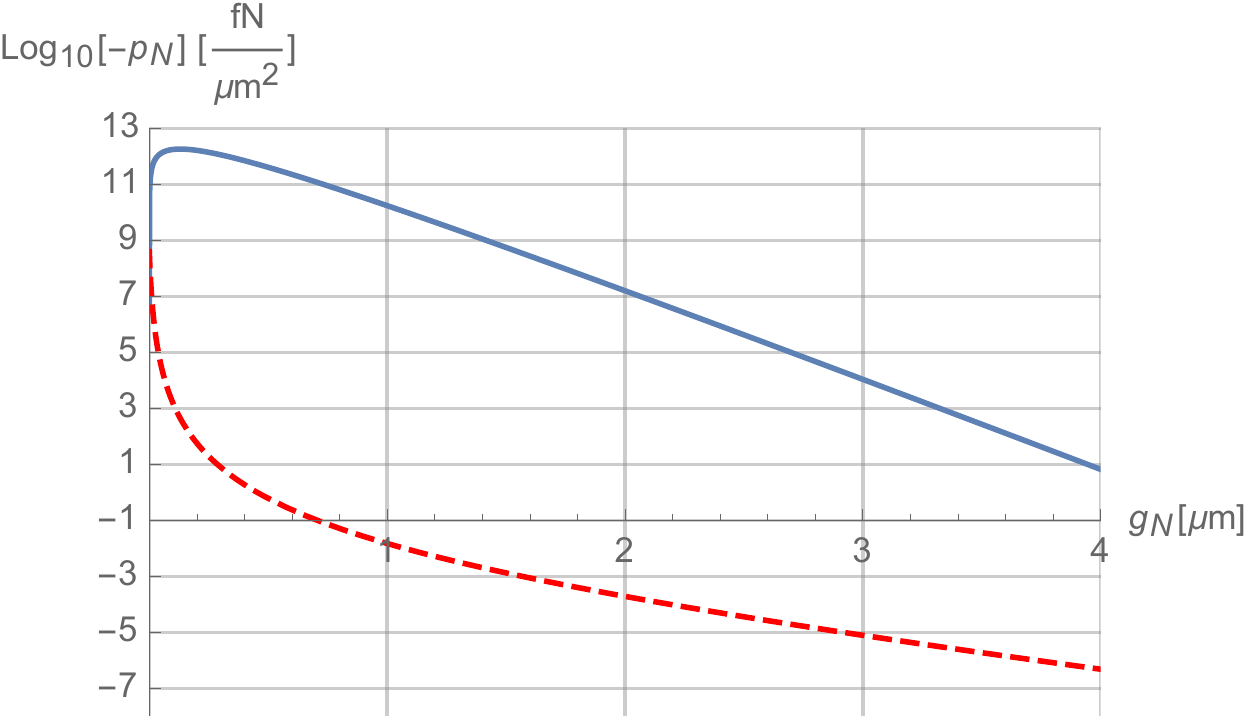}
\caption{According to  \cite{RonanEtAlBMM2014}  (continuum) compared to  \cite{GolestanehBMM2016} (dashed)  }
\end{subfigure}
\caption{Comparison between attractive forces. }
\label{fig:Experiment-4}
\end{figure}

The minimum concentration $\rho_{rl}(\vect{x})$ selected in \cite{GolestanehBMM2016} was quite high ($10^{5}$ receptors per $\mu {\rm m}^2$) compared to the concentrations of species that have been measured in \cite{DamioliEtAlSR2017}. 
Note also that the term $\rho_{rl}(\vect{x})$ should not be considered as constant, unless it refers to {\em{all receptors on the membrane}}, which seems illogical. Therefore, although the maximum number of moles or molecules per unit area for any species remains unchanged in time in the current configuration as stated in \eqref{eq:caRmax}, the transport processes affect the amount $\rho_{rl}(\vect{x})$ and {\em{induce a strong coupling between mechanical processes in the bulk and chemo-transport processes on the membrane}}.
Giving these numbers for granted, the resulting behavior of the Neumann electrostatic attractive tractions is plotted in Fig. \ref{fig:Experiment-4}.

According to equation \eqref{eq:attractiveforces}, attractive forces are inversely proportional to the distance between receptors and ligands, and those forces are infinitely high at contact. To get rid of this paradoxical statement, a strictly positive lower bound $h_0$ shall be defined together with a gap $g_N = h - h_0$ with $h_0$ being the gap between the cell and the substrate at contact; authors in \cite{GolestanehBMM2016} suggest $h_0 = 9.0 \times 10^{-3} \mu{\rm m}$. Repulsive forces are expected for distances below such a bound, as in Lennard-Jones potentials, yet this is not the case of equation \eqref{eq:attractiveforces}. 

On account of the values provided in \cite{GolestanehBMM2016}, accepting also the questionably high concentration $\rho_{rl}(\vect{x})$ that has been selected therein, attractive forces turn out to be remarkably high at $h_0$, as they concern integrins binding forces. Nonetheless, attractive forces decay rapidly and at a distance of ${\rm0.5} \mu{\rm m}$ they amount to a few $\rm fN / \mu m^2$. 

Numerical simulations, to appear in a companion publication, show also that those attractive forces, their range being so short, are not able to cause the cell spreading unless the characteristic size of the latter becomes very small. Indeed, authors in \cite{GolestanehBMM2016} considered a cell with radius ($\rm 12.5 nm$)  three orders of magnitude smaller than the measured radius of an endothelial cell in suspension (about $\rm 10 {\mu}m$). Size effects in mechanobiology are well known, and we argue that the role of attractive forces in virus receptors mediated endocytosis that has been pointed out \cite{Gao2016} and in nano-scale cells studied in \cite{GolestanehBMM2016}  does not apply to  the spreading of a micron-size endothelial cell. 
This remark is somewhat confirmed by analyzing the attractive forces used in \cite{RonanEtAlBMM2014}, namely 
\begin{equation}
\label{eq:attractiveforcesVikram}
\vect{t}(\vect{x}) = - Q  \frac{g_N}{\delta_p} \, \exp(- \frac{g_N}{\delta_p}  ) \,  \ey
\end{equation}
with $Q, \delta_p$ calibrated as $\rm 50 kPa = 5*10^7 fN/{\mu m}^2$ and $\rm 0.13 \mu m$, respectively.
They turn out to be 4 orders of magnitude higher than \eqref{eq:attractiveforces}, in order to allow cell spreading. 
We could not find justification in the literature for such a huge value of the ligand-receptor binding force acting on such a long range extent, hence we argue again that interaction forces of electrostatic nature are directly responsible of long range attraction and of spreading. {\em{Rather, these interactions are followed by the extension of pseudopodia from the cell body. As the cell begins to flatten against the substrate, it forms additional bonds, rearranges its cytoskeleton to form actin filaments and bundles, creating new focal adhesions. 
%
%
Spreading thus is a result of extensional and contractile forces exerted by pseudopodia and the cytoskeleton machinery \cite{Reinhart-King2005}. }}

\section{Thermodynamics }
\label{sec:thermodynamics}

The quest of the right thermodynamic principles in mechanobiology is, on one hand, far from being understood and, from a wider perspective, it paves the way to boundless questions of philosophical and ethical nature, 
as for the establishment of a thermodynamics of life \cite{WhatisLife}, which fall completely out of the scope of present paper. Major accomplishments have been recently achieved \cite{ShishvanBMM2018} in formulating fresh concepts that deviate from classical results of thermodynamics of non equilibrium. In this scientific area, which is nowadays flourishing, new fundamentals assertions are expected in the years to come.

Being aware of these deficiencies, we admit that our formulation of non equilibrium thermodynamics \cite{deGrootBook,SalvadoriEtAlJMPS2018} may not be able to capture some principles of mechanobiology that rule the dynamic of receptors - as for the homeostatic constraint - and we are prone to deepen our formulation in future studies.

\subsection{Thermodynamics of receptors motion on the membrane}

\subsubsection{Energy Balance}

As in section \ref{sec:Definitions}, denote with $\Omega(t)$ the advecting cell, and with $\partial \Omega(t)$ its lipid membrane. Consider an arbitrary region $ {\cal P}(t) \subset \, \partial \Omega(t) $. The first law of thermodynamics represents the balance of the interplay among the internal energy of $ {\cal P}(t) $, the heat transferred in $ {\cal P}(t) $ and the power due to mass exchanged by receptor dynamics on $ {\cal P}(t) $. The energy balance for the problem at hand reads: 
\begin{equation}
\label{eq:globalenergybalance}
\frac{ {\rm d} \, {\cal U}}{{\rm d} t} ({{\cal P}}) = {\cal Q}_u({{\cal P}}) + {\cal T}_u({ {\cal P}}) \; ,
\end{equation}
where $ {\cal Q}_u $ is the power due to heat transfer and ${\cal T}_u $ is the power due to mass transfer. Denoting with $ \partial {\cal P}(t)$ the bounding closed curve of $ {\cal P}(t)$ (see Fig. \ref{fig:Notation}), they read:
\begin{subequations} \label{eq:globalenergybalanceterms}
\begin{align}
{\cal Q}_u &= \int_{\cal P} s_q  \, {\rm d}a - \oint_{ \partial {\cal P} } \vect{q} \cdot \vect{t}_\perp  \, {\rm d} \ell  \\
\nonumber 
{\cal T}_u & = \int_{\cal P}  {\mu_{L}^{u}} \, s_L +  {\mu_{R}^{u}} \, s_R  \, {\rm d}a - \oint_{ \partial {\cal P} }{\mu_{R}^{u}} \, \vect{h}_R \cdot \vect{t}_\perp  \, {\rm d} \ell 
\end{align} 
\end{subequations}
The time variation of net internal energy $ {\cal U} $ thus corresponds to the power expenditure of two external agents: a heat contribution $ {\cal Q}_u $ where $ s_q $ is the heat supplied by external agents and $ \vect{q} $ is the heat flux vector; a mass contribution  {${\cal T}_u$}  in which the scalar ${\uchempot{\beta}}$ denotes the {change in specific energy provided by a unit supply of {\it{moles}}} of species $\beta = L, R$. Mass supply $s_L$ is the push-forward of the ligand supply $s_{L_R}({\vect{X}}, t)$ defined in eq. \eqref{eq:mass_balance_ref_L_final} and $\vect{h}_R $ is the flux of receptors along the membrane in the current configuration.

The net internal energy can be denoted in terms of specific internal energy $ u $ per unit surface, namely:
\begin{equation}
{\cal U}({\cal P}) = \int_{\cal P} u \, {\rm d}a \, .
\end{equation}
Applying the surface divergence theorem \eqref{eq:DivergenceTheorem} and mass balances leads from (\ref{eq:globalenergybalanceterms}) to
\begin{subequations} \label{eq:globalenergybalanceterms2}
	\begin{align}
	{\cal Q}_u= \int_{\cal P} s_q  \, {\rm d}a - \int_{  {\cal P} } \surfacedivergence{\vect{q}}{\cal P}   \, {\rm d}a  \; \, \qquad 
	{\cal T}_u = \int_{\cal P} {\mu_{L}^{u}} \, s_L +  {\mu_{R}^{u}} \, s_R  \, {\rm d}a - \int_{{\cal P} } \surfacedivergence{{\mu_{R}^{u}} \, \vect{h}_R}{\cal P} \, {\rm d}a  
	\; ,
	\end{align} 
\end{subequations}
whence the first law of thermodynamics arises\footnote{Since it must hold for any region $ {\cal P}(t) $, the current configuration local form of the first principle can be derived exploiting Reynold's theorem \eqref{eq:ReynoldsAdvectingSurface} on ${\cal P}(t)$
\begin{equation} \label{eq:energybalance}
\frac{ \partial u}{ \partial  t}
\, 
+
 \; \surfacedivergence{ u \,  \vect{v}_{adv} }{\cal P}
=
  s_q -\surfacedivergence{\vect{q}}{\cal P}  + {\mu_{R}^{u}} \frac{\partial {c_R}}{\partial t} + {\mu_{L}^{u}} \frac{\partial {c_L}}{\partial t}+ {\mu_{C}^{u}} \frac{\partial{c_C}}{\partial t} - \vect{h}_R \cdot \surfacegradient {\mu_{R}^{u}}{\cal P} 
+ \left( {\mu_{R}^{u}}+{\mu_{L}^{u}} - {\mu_{C}^{u}} \right)  w^{(\ref{eq:chem_react})} 
\; .
\end{equation} 
}
\begin{equation*}
\frac{ {\rm d} }{{\rm d} t}  \int_{\cal P}  u \, {\rm d}a = \int_{\cal P} s_q  \, -  \surfacedivergence{\vect{q}}{\cal P}  \, - \surfacedivergence{{\mu_{R}^{u}} \, \vect{h}_R}{\cal P} \, + {\mu_{L}^{u}} \, s_L  +  {\mu_{R}^{u}} \, s_R \, {\rm d}a  
\; .
\end{equation*}
It can be pulled back to the reference configuration in view of definitions of reference molarity of species ${c_{a_R}}( {\vect{X}}, t)$ in eq. \eqref{eq:caR}, of the reference flux vector ${ \vect{h}}_{a  _R}( {\vect{X}}, t)$ and of the reference mass supply $s_{a  _R}( {\vect{X}}, t)$  in eq. \eqref{eq:haR}, which readily extends to heat fluxes and supplies
\begin{equation}
\frac{ {\rm d} }{{\rm d} t}  \int_{{\cal P}_R}  u_R \, {\rm d}A 
= 
\int_{{\cal P}_R}  \, s_{q_R}  \, -  \surfaceDivergence{\vect{q_R}}{{\cal P}_R}  \, - \surfaceDivergence{{\mu_{R_R}^{u}} \, \vect{h}_{R_R}}{{\cal P}_R} \, + {\mu_{L_R}^{u}} \, s_{L_R}  +  {\mu_{R_R}^{u}} \, s_{R_R} \, {\rm d}A  
\; .
\end{equation}
Since it must hold for any region $ {\cal P}_R $, the local form of the first principle can be derived exploiting the mass balance equations  \eqref{eq:mass_balance_ref_R}, \eqref{eq:mass_balance_ref_C}, \eqref{eq:mass_balance_ref_L_final} 
in the reference configuration
\begin{align} 
\label{eq:ref_energybalance}
\nonumber
\frac{ {\rm d} u_R}{ {\rm d}  t}
\, 
&
=
  s_{q_R} -\surfaceDivergence{\vect{q_R}}{{\cal P}_R}  
  - \vect{h}_{R_R} \cdot \surfaceGradient {\mu_{R_R}^{u}}{{\cal P}_R} 
\\
&  
  + {\mu_{R_R}^{u}} \, \frac{\partial {c_{R_R}}}{\partial t} 
  + {\mu_{L_R}^{u}} \,  \frac{\partial {c_{L_R}}}{\partial t}
  + {\mu_{C_R}^{u}} \,  \frac{\partial{c_{C_R}}}{\partial t} 
  + \left( {\mu_{R_R}^{u}}+{\mu_{L_R}^{u}} - {\mu_{C_R}^{u}} \right)  w_R^{(\ref{eq:chem_react})}  
\; .
\end{align}

\subsubsection{Entropy balance equations}

The second law of thermodynamics represents the balance of the interplay among the internal entropy of $ {\cal P} $ and the entropy transferred in $ {\cal P} $ due to mass exchange and heat transferred on $ {\cal P} $. The entropy balance for the problem at hand reads:

\begin{equation}
\label{eq:globalentropybalance}
	\frac{{\rm d}  S}{{\rm{d}} t} ({\cal P})\, - \, 	\frac{{\rm{d}}  S_{irr}}{{\rm{d}} t}({\cal P}) = \, {\cal Q}_\eta({{\cal P}}) + {\cal T}_\eta({{\cal P}}) \; ,
\end{equation}
\noindent where $ S $ is the net internal entropy of $ {\cal P} $, $ S_{irr} $ is the entropy produced inside $ {\cal P} $, $ Q_\eta $ the entropy per unit time due to heat transfer, $ T_\eta $ the entropy per unit time due to mass transfer. The individual contributions read:
\begin{subequations} \label{eq:globalentropybalanceterms}
	\begin{align}
	Q_\eta & =  \int_{ {\cal P} } \frac{s_q}{T} \, {\rm d}A \, - \,\oint_{ \partial {\cal P} } \frac{\vect{q}}{T} \cdot  \vect{t}_\perp  \, {\rm d} \ell 	
	\; , \\
	T_\eta & =  \int_{\cal P} {\mu_{L}^{\eta}} \, s_L +  {\mu_{R}^{\eta}} \, s_R  \, {\rm d}A - \oint_{ \partial {\cal P} } \mu_{R}^{\eta} \, \vect{h}_R  \cdot  \vect{t}_\perp  \, {\rm d} \ell 
	\; .
	\end{align} 
\end{subequations}
The scalar $ \mu_{\beta}^{\eta} $ denotes the change in specific entropy provided by a unit supply of moles of species $ \beta $. 
Equation $ \eqref{eq:globalentropybalance} $ stems from the non-trivial assumption that mechanics does not contribute directly to the total entropy flow in the entropy balance equation \cite{SalvadoriEtAlJMPS2018}.
The second law of thermodynamics states that:
\begin{equation}
\frac{{\rm{d}}  S_{irr}}{{\rm{d}} t} \geq 0.
\end{equation}
Analogously to the  energy counterpart, we define the specific internal entropy $ \eta $ per unit volume and write the entropy imbalance in the reference configuration as
%
\begin{equation*}
\frac{ {\rm d} }{{\rm d} t}
\int_{{\cal P}_R}  \, \eta_R \, {\rm d}A   
+  
\int_{{\cal P}_R} - \frac{s_{q_R}}{T} + \surfaceDivergence{\frac{\vect{{q_R}}}{T}}{{\cal P}_R}  \, - \,{\mu_{L_R}^{\eta}} \, s_{L_R}  - {\mu_{R_R}^{\eta}} \, s_{R_R} +  \surfaceDivergence{\mu_{R_R}^{\eta} \, \vect{h}_{R_R} }{{\cal P}_R}  \,  \, {\rm d}A \, \geq 0
\; .
\end{equation*}
After multiplication by $ T \geq 0 $, replacing $ - s_{q_R} + \,  \surfaceDivergence{\vect{q_R}}{{\cal P}_R} $ by means of the energy balance \eqref{eq:ref_energybalance},  and some simple algebra, the local form of the entropy imbalance becomes
\begin{equation}
\begin{aligned}
T\,  \frac{{\rm d} \eta_R }{{\rm d} t} 
&
 - \frac{{\rm d} u_R}{{\rm d} t} 
 + \, \frac{\partial {c_{R_R}}}{\partial t} \, \left[ {\mu_{R_R}^{u}} - T  \, {\mu_{R_R}^{\eta}} \right]  
 + \, \frac{\partial {c_{L_R}}}{\partial t} \, \left[ {\mu_{L_R}^{u}} - T  {\mu_{L_R}^{\eta}} \right] 
 + \, \frac{\partial {c_{C_R}}}{\partial t} \left[ {\mu_{C_R}^{u}}  - T  {\mu_{C_R}^{\eta}} \right]   + 
 \\ 
& 
 - \frac{1}{T} \, \vect{q_R} \cdot \surfaceGradient{ T }{{\cal P}_R} 
+ T \, \vect{h}_{R_R} \cdot \surfaceGradient {\mu_{R_R}^{\eta}}{{\cal P}_R}   
 - \vect{h}_{R_R} \cdot \surfaceGradient{\mu_{R_R}^{u}}{{\cal P}_R} 
 \\ 
& 
 + \left( {\mu_{R_R}^{u}} - T  \, {\mu_{R_R}^{\eta}}+ {\mu_{L_R}^{u}} - \, T  {\mu_{L_R}^{\eta}} - \mu_{C_R}^{u} + \, T  {\mu_{C_R}^{\eta}} \right) w_R^{(\ref{eq:chem_react})} 
 \geq 
 0
\end{aligned}
\end{equation}
Denote with $\beta=R,L,C$ and with the symbols $ \mu_{\beta_R} $, $ A_R^{\eqref{eq:chem_react}} $ the quantities
\begin{equation}
\label{eq:chempotential}
\mu_{\beta_R} = \mu_{\beta_R}^{u} - T \, \mu_{\beta_R}^{\eta}
\end{equation}
\begin{equation}
\label{eq:aff}
A_R^{\eqref{eq:chem_react}}  = -\mu_{R_R} -\mu_{L_R}+ \mu_{C_R}
\; .
\end{equation}
By noting that:
\begin{equation*}
	T \, \vect{h}_{R_R} \cdot \surfaceGradient {\mu_{R_R}^{\eta}}{{\cal P}_R} = \vect{h}_{R_R} \cdot \surfaceGradient{ T \, {\mu_{R_R}^{\eta}}  }{{\cal P}_R} -  \vect{h}_{R_R} \cdot \surfaceGradient {T}{{\cal P}_R} \, {\mu_{R_R}^{\eta}}
\end{equation*}
one finally writes the entropy imbalance as:
\begin{equation}
\label{eq:ref_entropybalance}
\begin{aligned}
T
\,  \frac{{\rm d} \eta_R }{{\rm d} t} 
&
 - \frac{{\rm d} u_R}{{\rm d} t} 
 + \, \frac{\partial {c_{R_R}}}{\partial t} \,  {\mu_{R_R}}    
 + \, \frac{\partial {c_{L_R}}}{\partial t} \,  {\mu_{L_R}}  
 + \, \frac{\partial {c_{C_R}}}{\partial t}  {\mu_{C_R}}    + 
 \\ 
& 
- \left( \frac{1}{T} \, \vect{q_R} + \, {\mu_{R_R}^{\eta}} \vect{h}_{R_R} \right) \cdot \surfaceGradient {T}{{\cal P}_R} 
 - \vect{h}_{R_R} \cdot \surfaceGradient{\mu_{R_R}}{{\cal P}_R} 
 \\ 
& 
 - A_R^{\eqref{eq:chem_react}} \, w_R^{(\ref{eq:chem_react})} 
 \geq 
 0
 \; .
\end{aligned}
\end{equation}

\subsubsection{Helmholtz Free Energy and thermodynamic restrictions} 

The referential specific Helmholtz free energy per unit volume is defined as:
\begin{equation}
\label{eq:HFE}
	\psi_R = u_R - T \, \eta_R
\end{equation}
and is taken as a function of temperature and concentrations,  $ \psi_R \left( T, c_{R_R}, c_{L_R}, c_{C_R} \right) $. It thus hold:
\begin{equation*}
\, \Temperature \,  \frac{ {\rm d} \eta_R }{{\rm d} t} 
				 - \, \frac{ {\rm d} u_R }{{\rm d} t}  = 
				 - \, \frac{ {\rm d} \psi_R }{{\rm d} t} \,-  \eta_R \,  \frac{ \partial \Temperature }{\partial t}  
 =  
 - \frac{\partial \psi_R }{\partial c_{L_R}} \frac{\partial c_{L_R} }{\partial t} - \frac{\partial \psi_R }{\partial c_{R_R}} \frac{\partial c_{R_R} }{\partial t} - \frac{\partial \psi_R }{\partial c_{C_R}} \frac{\partial c_{C_R} }{\partial t} - \left( \eta_R + \frac{\partial \psi_R}{\partial T}   \right)  \frac{\partial T }{\partial t}
\end{equation*}
which can be plugged in $ \eqref{eq:ref_entropybalance} $ to derive the entropy imbalance in the Clausius-Duhem form:
\begin{equation} 
\label{eq:clausius:inequality}
\begin{aligned}
& 
\left( - \frac{\partial \psi_R }{\partial c_{R_R}} + \mu_{R_R} \right) \frac{\partial {c_{R_R}}}{\partial t}  +\left( - \frac{\partial \psi_R }{\partial c_{L_R}} + \mu_{L_R} \right) \frac{\partial {c_{L_R}}}{\partial t} + \left( - \frac{\partial \psi_R }{\partial c_{C_R}} + \mu_{C_R} \right) \frac{\partial {c_{C_R}}}{\partial t} - \left( \eta_R + \frac{\partial \psi_R }{\partial T} \right) \frac{\partial T }{\partial t}  + \\
& \qquad  -\frac{1}{T} \vect{\underline{q_R}} \cdot \surfaceGradient{ T}{{\cal P}_R}  - A_R^{(\ref{eq:chem_react})}  w_R^{(\ref{eq:chem_react})} - \vect{h}_{R_R} \cdot \surfaceGradient{ \mu_{R_R} }{{\cal P}_R}  \, \geq 0
\end{aligned}
\end{equation}
with $\vect{\underline{q_R}} = \vect{q_R} + \, T\, {\mu_{R_R}^{\eta}} \vect{h}_{R_R}$.
This inequality must hold for any value of the time derivative of the temperature and of the referential concentrations $ c_{R_R}$, $c_{L_R} $, and $ c_{C_R} $. Since they appear linearly in the inequality, the factors multiplying them must
be zero, as otherwise it would be possible to find a value for the time derivatives that violate the inequality. Therefore, the following restrictions apply
\begin{equation}
\label{eq:TDequalities}
	\mu_{R_R}= \frac{\partial \psi_R }{\partial c_{R_R}}, \qquad  
	\mu_{L_R}= \frac{\partial \psi_R }{\partial c_{L_R}} , \qquad 
	\mu_{C_R}= \frac{\partial \psi_R }{\partial c_{C_R}}, \qquad 
	\eta_R= -\frac{\partial \psi_R }{\partial T} 
	\;
	.
\end{equation}

In view of formula \eqref{eq:TDequalities}, the amount $\mu_\beta$ declared in eq. \eqref{eq:chempotential} acquires the meaning of  {\it{chemical potential}} and hence the term $A^{(\ref{eq:chem_react})} $ in eq. \eqref{eq:aff} turns out to be the {\it{affinity of the reaction}} (\ref{eq:chem_react}).  Further remarks on this thermodynamic approach can be found in \cite{SalvadoriEtAlJMPS2018}.

\bigskip

Equation \eqref{eq:TDequalities} yields to the so called Clausius-Plank inequality: 
\begin{equation} 
\label{eq:CP_inequality}
 \qquad  -\frac{1}{T} \vect{\underline{q_R}} \cdot \surfaceGradient{ T}{{\cal P}_R}  - A_R^{(\ref{eq:chem_react})}  w_R^{(\ref{eq:chem_react})} - \vect{h}_{R_R} \cdot \surfaceGradient{ \mu_{R_R} }{{\cal P}_R}  \, \geq 0
\end{equation}
that splits under the assumptions of Curie's principle 
in the following set of inequalities:
\begin{subequations}
 \label{eq:curie}
\begin{align}
& 
\frac{1}{T} \vect{\underline{q_R}} \cdot \surfaceGradient{ T}{{\cal P}_R}  +
 \vect{h}_{R_R} \cdot \surfaceGradient{ \mu_{R_R} }{{\cal P}_R} 
 \leq 0  \; ,
\\
& A_R^{(\ref{eq:chem_react})}  \, w_R^{(\ref{eq:chem_react})}  \leq 0  \; .
\end{align}
\end{subequations}

\subsubsection{Constitutive theory}

We will assume henceforth that the lipid membrane is in thermal equilibrium, i.e. $\surfaceGradient{ T}{{\cal P}_R}  =\vect{0} $, and that the Helmholtz free energy density is additively decomposed into three separate parts:
\begin{equation}
\psi_R \left( c_{R_R}, c_{L_R}, c_{C_R} \right) = \psi_R^R (c_{R_R}) + \psi_R^L (c_{L_R}) +\psi_R^C (c_{C_R}) 
\end{equation}
meaning that the contributions of species are uncoupled, neglecting molecular friction that would lead to a Maxwell-Stefan description of transport.
The free energy density of mobile guest atoms
interacting with a host medium is described by an ideal solution model, which stems from a statistical mechanics description of the entropy for isolated systems in terms of the density of states, i.e. the number of possible molecular configurations \cite{ShellBook2015} in the case of two-state systems.  Making recourse to Stirling's approximation, one finds that the formula for combinations provides the following free energy density for the continuum approximation of mixing \cite{ShellBook2015} of the generic species $\beta = R,L,C$ 
\begin{equation}
\psi_R^\beta(c_{\beta_R})= \mu_{\beta_R}^{0} \, c_{\beta_R} + R \, T c_{\beta_R}^{max} \left[ \vartheta_{\beta_R} \ln \vartheta_{\beta_R} + (1- \vartheta_{\beta_R} ) \ln (1- \vartheta_{\beta_R} )\right] 
\; ,
\end{equation}
with $\vartheta_{\beta_R} $ defined in \eqref{eq:varthetaRmax} as the ratio between the concentration and the saturation limit for each species in the reference configuration. The chemical potential descends from eq. \eqref{eq:TDequalities}
\begin{equation}
\label{eq:mubeta}
\mu_{\beta_R} = \frac{\partial \psi_R^\beta}{\partial c_{\beta_R}} = \mu_{\beta_R}^{0} + R \, T \left( \ln \vartheta_{\beta_R} - \ln \left(1-\vartheta_{\beta_R}  \right) \right) 
\; .
\end{equation}

\bigskip

\noindent A strategy to meet the thermodynamic restriction (\ref{eq:curie}a) is to model the flux of receptors by Fickian-diffusion, that linearly correlates $\vect{h}_{R_R} $ to the gradient of its chemical potential $ {\mu}_{R_R} $:
\begin{equation}
\label{eq:Ficksalpha}
\vect{h}_{R_R} = - \tensor{M}_{R}(c_R) \; \surfaceGradient{  {\mu}_{R_R} }{{\cal P}_R} 
\end{equation}

\noindent by means of a positive definite mobility tensor $\tensor{M}_R $. 
The following isotropic non linear specialization for the mobility tensor $\tensor{M}_{R} $ is chosen \cite{AnandJMPS2012}
\begin{equation}
\label{eq:isotropicmobility1}
    \tensor{M}_{R} ( c_{R_R} ) = \mobility_{R} \, c_{R_R}^{max} \; \vartheta_{R_R} \, \left( 1 -  \vartheta_{R_R} \right)\; \mathds{1} \; ,
\end{equation}
where $c_{R_R}^{max} $ is the saturation limit for receptors, and $\mobility_{R}>0 $  is
the {\it{mobility}} of receptors. Definition (\ref{eq:isotropicmobility1}) represents the physical requirement that both the pure ($c_{R_R}=0$) and the saturated ($ c_{ R_R } = c_{R_R}^{max} $) phases have vanishing mobilities.  Neither the mobility $\mobility_R$ nor the saturation concentration $c_{R_R}^{max}$ are assumed to change in time. 
Whereby experimental data indicate an influence of temperature, stresses, or concentrations, such a limitation can be removed without altering the conceptual picture.
Noting that
\begin{equation*}
\surfaceGradient{  {\mu}_{R_R} }{{\cal P}_R}  =  \frac{ R \, T }{c_{R_R}^{max}} \, \frac{1}{\vartheta_{R_R} (1- \vartheta_{R_R})} \; \surfaceGradient{ c_{R_R} }{{\cal P}_R}  
\; ,
\end{equation*}
Fick's Law \eqref{eq:Ficksalpha} specializes as 
\begin{equation} \label{flux:react1}
\vect{h}_{R_R} = - \diffusivity_{R} \,  \surfaceGradient{ c_{R_R} }{{\cal P}_R} 
\; ,
\end{equation}
where $\diffusivity_{R} = \mobility_{R} \, R \, \Temperature$ is the receptor {\it{diffusivity}}.

\subsubsection{Chemical kinetics}
\label{subsec:chemkin}

The chemical kinetics of reaction $ \eqref{eq:chem_react} $ is modeled via the law of mass action \eqref{eq:mass_action_ref}. 
%
%
%
%
%
Experimental evidences \cite{DamioliEtAlSR2017} show that: (i) the equilibrium constant \eqref{eq:eq_const} is high, thus favoring the formation of ligand-receptor complex and the depletions of receptors and ligands; (ii) the diffusion of receptors on the cell membrane is much slower than interaction kinetics. Accordingly,
it can be assumed that the reaction kinetics is infinitely fast, in the sense that the time required to reach chemical equilibrium is orders of magnitude smaller than the time-scale of other processes. 
For these reasons we assume that the concentrations of species are ruled by thermodynamic equilibrium at all times, and the concentration of complex ${c_{C_R}}$ is related to the others by the equation \eqref{cond:ref:concentration1}. This very same equation could be derived imposing 
$$ A^{(\ref{eq:chem_react})}=0 \; .$$ Simple algebra allows deriving eq. \eqref{cond:ref:concentration1}, provided that to the equilibrium constant $K_{\rm eq}^{(\ref{eq:chem_react})} $ the alternative definition
\begin{equation}
\label{eq:Keq}
K_{\rm eq}^{(\ref{eq:chem_react})}  =  \, \exp \left(  - \frac{ \Delta G^0}{R \, T} \right) 
\end{equation}
is given, where $ \Delta G^0= \mu_{C}^{0} - \mu_{L}^{0} - \mu_{R}^{0} $ is the standard Gibbs free energy.

\subsection{Thermo-chemo-mechanics of cells}

Endothelial cells show two main paradigmatic mechanical attitudes: active and passive. Active response is related to the ability of the cell to change, as a result of external cues,
its own cytoskeletal conformation, i.e. to reorganize the morphology of the biopolymers net that provides the structural resistance during adhesion (to the ECM or to other cells), migration (e.g. chemotaxis, mechanotaxis, and durotaxis) and division (eg. mitosis).  Passive, instead, refers to the mechanical response that each component of the cell has inasmuch material bodies, 
in accordance with their own internal structure and as a result of external actions.

\subsubsection{Energy balance}
\label{subsec:firstprinciple}

%
Define in the bulk an arbitrary region ${Q}(t) \subset \Omega(t)$. 
The energy balance for the problem at hand, using the notation introduced in \cite{SalvadoriEtAlJMPS2018}, reads:
\begin{equation}
\label{eq:bulk:globalenergybalance}
\frac{ {\rm d} \, {\cal U}}{{\rm d} t} ({Q})   = {\cal W}_u({Q}) + {\cal Q}_u({Q}) + {\cal T}_u({Q}) 
  \; ,
\end{equation}
with $\cal U$ the net internal energy of ${Q}$, ${\cal W}_u$ the mechanical external power, ${\cal Q}_u$ the power due to heat transfer, ${\cal T}_u$ the power due to mass exchanged by actin dynamics on $ {Q}(t) $. It is assumed that each of these processes is {\em{energetically separable}} in the balance. The individual contributions read:
\begin{subequations}
\begin{eqnarray}
{\cal W}_u({Q}) &=& \int_{Q}  \vect{b} \cdot \vect{v} \;  {\rm d} \Omega + \int_{\partial Q} \, \vect{t} \cdot \vect{v} \;  {\rm d} \Gamma
  \; ,
\\
{\cal Q}_u({Q})  &=& \int_{Q} \, s_q \,  {\rm d} \Omega - \int_{\partial Q} \, \vect{q} \cdot \vect{n} \; {\rm d} \Gamma
 \; ,
\\
{\cal T}_u({Q})  &=&  \int_{Q}  \,  \uchempot{G} \, s_G \, +  \,  \uchempot{F} \, s_{F} \; {\rm d} \Omega - \int_{\partial Q}  \,  \uchempot{G} \, \vect{h}_G \cdot \vect{n} \, {\rm d} \Gamma  
  \; .
\end{eqnarray}
\label{eq:bulk:globalenergybalanceterms}
\end{subequations}
Assumption \eqref{eq:zero:filament_flux} has been accounted for in the mass transfer contribution ${\cal T}_u({Q})$.

The time variation of net internal energy $\cal U$ corresponds to the power expenditure of external agencies: 
a mechanical contribution {${\cal W}_u$} due to body forces $\vect b$ and surface tractions $\vect{t}$ that do work on velocities $ \vect{v} $; 
a heat contribution  {${\cal Q}_u$}  where $s_q$ is the heat supplied by external agencies and $\vect{q}$ is the heat flux vector; a mass contribution  {${\cal T}_u$}  in which the scalar $\uchempot{\beta}$ denotes the {change in specific energy provided by a unit supply of {\it{moles}}} of $\beta = G, F$ actin. Mass supply $s_\beta$ is the push-forward of the supply $s_{\beta_R}({\vect{X}}, t)$ defined in eq. \eqref{eq:JhaR} and $\vect{h}_G $ is the flux of G-actin in the current configuration.

\bigskip
Standard application of the divergence theorem and of balance equations leads from (\ref{eq:bulk:globalenergybalanceterms}a) to
\begin{eqnarray}
{\cal W}_u({Q}) &=& \int_{Q}  \,   \tensor{ \sigma} :  {\tensor{l} }
\,  {\rm d} \Omega 
\; .
\label{eq:bulk:globalenergybalanceterms2}
\end{eqnarray}
where 
$ {\tensor{l} }$ is the gradient of velocity tensor, i.e. $ {\tensor{l} } = { \gradient{ \vect{ v }}}$ and 
$\tensor{ \sigma} $ is the Cauchy stress tensor. 
Since it is well known that
$$ 
  \tensor{ \sigma} :  {\tensor{l} }\,  {\rm d} \Omega 
  =
  \tensor{ P} :  {\dot{\tensor{F}} }\,  {\rm d} \Omega_R   
  =
  \tensor{ S} :  {\dot{\tensor{E}} }\,  {\rm d} \Omega_R   
  \; ,
$$
the mechanical power expenditure can be written in terms of the first Piola-Kirchoff stress $\tensor{ P} = J  \, \tensor{ \sigma} \, \tensor{F}^{-T}$ or of the second Piola-Kirchoff stress $\tensor{ S} = \tensor{F}^{-1} \tensor{P}$ in the referential configuration.
Analogously, by defining the referential heat flux $ \vect{ q}_R = J \, \tensor{F}^{-1} \, \vect{ q}$ and making use of Nanson's formula, it holds
\begin{equation} 
\label{eq:bulk:referential_flux}
  \vect{q} \cdot \vect{n} \; {\rm d} \Gamma
  =
  \vect{q}_R \cdot \vect{n}_R \; {\rm d} \Gamma_R
  \; .
\end{equation} 
As usual in the thermodynamics of continua, see e.g. \cite{GurtinFriedAnand}, one can make use of the specific internal energy $u_R$ per unit volume in the reference configuration 
to write the referential local form of the first principle as
\begin{align}
\label{eq:bulk:intenlocalform1}
  \frac{ {\rm d}  u_R}{{\rm d}  t}   
&
=
 \tensor{ S} :  {\dot{\tensor{E}} }\ + s_{q_R} - \Divergence{ \vect{q}_R}
  - \vect{h}_{G_R} \cdot \Gradient {\mu_{G_R}^{u}}
\\
&  
  + {\mu_{G_R}^{u}} \,  \frac{\partial {c_{G_R}}}{\partial t}
  + {\mu_{F_R}^{u}} \,  \frac{\partial{c_{F_R}}}{\partial t} 
  + \left( {\mu_{G_R}^{u}} - {\mu_{F_R}^{u}} \right)  w_R^{(\ref{eq:actin_polymerization})}  
                         \; .
\end{align}

\subsubsection{Entropy imbalance}
\label{subsec:secondprinciple}

The second law of thermodynamics represents the balance of the interplay among the internal entropy of $Q$ and the entropy transferred in it due to mass exchange and heat transferred. We make the non-trivial assumption that mechanics does not contribute directly to the total entropy flow in the entropy balance equation, as profoundly elaborated in \cite{deGrootBook, HolzapfelBook}.   The entropy balance for the problem at hand reads:
\begin{equation}
\label{eq:bulk:globalentropybalance}
\frac{ {\rm d} \, {\cal S}}{{\rm d} t} ({Q})  - \frac{ {\rm d} \, {\cal S}_i}{{\rm d} t} ({Q})   =  {\cal Q}_\eta({Q}) +  {\cal T}_\eta({Q}) 
  \; ,
\end{equation}
where $\cal S$ is the net internal entropy of ${Q}$, ${\cal S}_i$ is the entropy produced inside ${Q}$, ${\cal Q}_\eta$ the entropy per unit time due to heat transfer, ${\cal T}_\eta$ the entropy per unit time due to mass transfer. The individual contributions read:
%
\begin{eqnarray}
{\cal Q}_\eta({Q})  
&=& 
\int_{Q} \,  \frac{s_q}{\Temperature}  \,  {\rm d} \Omega - \int_{\partial Q} \, \frac{ \vect{q} }{\Temperature} \cdot \vect{n} \; {\rm d} \Gamma
  \; ,
\\
{\cal T}_\eta(Q)  &=&  \int_{Q}  \,  \etachempot{G} \, s_G \, +  \,  \etachempot{F} \, s_{F} \; {\rm d} \Omega - \int_{\partial Q}  \,  \etachempot{G} \, \vect{h}_G \cdot \vect{n} \, {\rm d} \Gamma  
%
  \; .
\label{eq:bulk:globalentropybalanceterms}
\end{eqnarray}
The second law of thermodynamics states that 
$$  \frac{ {\rm d} \, {\cal S}_i}{{\rm d} t} ({Q}) \ge 0 \; .$$

\bigskip
As for the energy, one can make use of the specific internal entropy $\eta_R$ per unit referential volume
to localize and 
rephrase the entropy imbalance in terms of internal energy  taking advantage of identity (\ref{eq:bulk:intenlocalform1}) and of the sign definiteness of temperature :
\begin{eqnarray}
\label{eq:bulk:entropylocalform2}
&&
   \Temperature \; \frac{ \rm d  }{{\rm d} t}    \; \eta_R  \, 
 -  \frac{ {\rm d}  u_R}{{\rm d}  t}   
 + \tensor{ S} :  {\dot{\tensor{E}} }\
 +   \frac{ \partial c_{G_R} }{\partial t}  \, \mu_{G_R} 
 +   \frac{ \partial c_{F_R} }{\partial t}   \, \mu_{F_R} 
 +
    			\\ &&
\nonumber
\qquad        
 -  \left( \frac{1}{T} \, \vect{q_R} + \, {\mu_{G_R}^{\eta}} \vect{h}_{G_R} \right)  \cdot \Gradient{\Temperature} 
			- \,  \vect{h}_{G_R} \cdot \Gradient{ \mu_{G_R} } 
                       	- A_R^{\eqref{eq:actin_polymerization}} \,  w_R^{(\ref{eq:actin_polymerization})}  
			\ge 0 
			  \; ,
\end{eqnarray}
having denoted with $\beta=G,F$ and with the symbols $ \mu_{\beta_R} $, $ A_R^{\eqref{eq:actin_polymerization}} $ the quantities
\begin{equation}
\label{eq:bulk:chempotential}
\mu_{\beta_R} = \mu_{\beta_R}^{u} - T \, \mu_{\beta_R}^{\eta}
\end{equation}
\begin{equation}
\label{eq:bulk:aff}
A_R^{\eqref{eq:actin_polymerization}}  =  -\mu_{G_R}+ \mu_{F_R}
\; .
\end{equation}

\subsubsection{Helmholtz free energy and thermodynamic restrictions} 

The referential specific Helmholtz free energy per unit volume $ \psi_R \left( T, c_{G_R}, c_{F_R}, \tensor {C}, \tensor{\xi} \right) $, defined as in \eqref{eq:HFE}, is taken as a function of temperature, strains (either $\tensor {C}$ or $\tensor {E}$), concentrations $c_{G_R}, c_{F_R}$, and of some kinematic internal variables $\tensor{\xi}$ that compare with the usual meaning in inelastic constitutive laws \cite{ GurtinFriedAnand, paolucci2016, HolzapfelBook, TadmoreBook1,SIMOCMAME1988a, SIMOCMAME1988b}.
It follows that
\begin{equation}
\label{eq:bulk:dotfreeen}
\, \Temperature \,  \frac{ {\rm d} \eta_R }{{\rm d} t} 
				 - \, \frac{ {\rm d} u_R }{{\rm d} t}  = 
				 - \, \frac{ {\rm d} \psi_R }{{\rm d} t} \,-  \eta_R \,  \frac{ \partial \Temperature }{\partial t}  
				 			\; ,
\end{equation}
which can be inserted in \eqref{eq:bulk:entropylocalform2} to derive the entropy imbalance in final form:
\begin{eqnarray}
&&
\label{eq:bulk:ColNoll3}			
			- \, \frac{ {\rm d} \psi_R }{{\rm d} t} \,-  \eta_R \,  \frac{ \partial \Temperature }{\partial t}  
 + \tensor{ S} :  {\dot{\tensor{E}} }\
 +   \frac{ \partial c_{G_R} }{\partial t}  \, \mu_{G_R} 
 +   \frac{ \partial c_{F_R} }{\partial t}   \, \mu_{F_R} 
 +
    			\\ &&
\nonumber
\qquad        
 -  \left( \frac{1}{T} \, \vect{q_R} + \, {\mu_{G_R}^{\eta}} \vect{h}_{G_R} \right)  \cdot \Gradient{\Temperature} 
			- \,  \vect{h}_{G_R} \cdot \Gradient{ \mu_{G_R} } 
                       	- A_R^{\eqref{eq:actin_polymerization}} \,  w_R^{(\ref{eq:actin_polymerization})}  
			\ge 0 
			  \; .
\end{eqnarray}

\bigskip
In view of the stated functional dependency of the free energy, its total derivative with respect to time reads:
\begin{equation}
\label{eq:bulk:intenrate}
 \frac{ {\rm d}   }{{\rm d}  t} { \psi_R ( \Temperature,  c_{G_R}, c_{F_R}, {\tensor {C}},  \tensor{\xi} ) } 
 = 
   \frac{\partial \psi_R}{\partial \Temperature} \, \frac{ \partial \Temperature }{\partial t}  + 
   \frac{\partial \psi_R}{\partial c_{G_R}} \, \frac{ \partial c_{G_R} }{\partial t}  + 
   \frac{\partial \psi_R}{\partial c_{F_R}} \, \frac{ \partial c_{F_R} }{\partial t}  + 
   \frac{\partial \psi_R}{ \partial \tensor {C} }  \, : \, \dot{ \tensor{C} }+
   \frac{\partial \psi_R}{\partial \tensor{\xi} } \, : \, \frac{ \partial \tensor{\xi} }{\partial t} 
\end{equation}
The Clausius-Duhem inequality yields:
\begin{eqnarray}
 &&
  \nonumber
  \left( - \frac{\partial \psi_R }{\partial c_{G_R}} + \mu_{G_R} \right) \frac{\partial {c_{G_R}}}{\partial t}  + 
  \left( - \frac{\partial \psi_R }{\partial c_{F_R}}  + \mu_{F_R} \right) \frac{\partial {c_{F_R}}}{\partial t} + 
    \,\frac{ \partial \Temperature }{\partial t}   
    \, \left( -\eta_R -  \frac{\partial \psi_R}{\partial \Temperature} \right) 
   + \; \dot{ \tensor{C} }: \left( \frac{1}{2} \, \tensor{ S}  -    \frac{\partial \psi_R}{ \partial \tensor {C} }  \right) 
   + \\
&& \qquad 
  			- \frac{\partial \psi_R}{\partial \tensor{\xi} } \, : \, \frac{ \partial \tensor{\xi} }{\partial t} 
 -  \left( \frac{1}{T} \, \vect{q_R} + \, {\mu_{G_R}^{\eta}} \vect{h}_{G_R} \right)  \cdot \Gradient{\Temperature} 
			- \,  \vect{h}_{G_R} \cdot \Gradient{ \mu_{G_R} } 
                       	- A_R^{\eqref{eq:actin_polymerization}} \,  w_R^{(\ref{eq:actin_polymerization})}  
			\ge 0 
			  \; .
\label{eq:bulk:entropylocalform2}
\end{eqnarray}
This inequality must hold for any value of the time derivative of the temperature, the referential concentrations, the strain tensor. Since they appear linearly in the inequality, the factors multiplying them must be zero, as otherwise it would be possible to find a value for the time derivatives that violate the inequality. Therefore, the following prescriptions apply
\begin{subequations}
\begin{equation}
	\tensor{ S}  =   2  \, \frac{\partial \psi_R}{ \partial \tensor {C} }  
\; , \qquad
    \eta_R  =  -   \frac{\partial \psi_R}{\partial \Temperature}   
  \; , \qquad
    \mu_{G_R}= \frac{\partial \psi_R }{\partial c_{G_R}}  
    \; , \qquad
    \mu_{F_R}= \frac{\partial \psi_R }{\partial c_{F_R}} 
\; .				 				 
\label{eq:bulk:TDequalities}
\end{equation}
The internal force, conjugate to $\tensor{\xi}$, will be denoted with the symbol $\tensor{\chi}$, i.e.
\begin{equation}
\label{eq:bulk:internalforces}
\tensor{\chi}_R = - \frac{\partial \psi_R }{\partial \tensor{\xi} }
\; .
\end{equation}
\label{eq:bulk:TDConsistency}
\end{subequations}
Equation \eqref{eq:bulk:TDequalities} yields to the Clausius-Plank inequality, which under the assumptions of Curie symmetry principle \cite{deGrootBook}, can be written as 
\begin{subequations}
\begin{eqnarray} 
\label{eq:bulk:TDindelsticdiss}
  &&
     \tensor{\chi}_R \, : \, \dot{ \tensor{\xi} }  
    \ge 0 
  \\ &&
  \label{eq:bulk:TDcrosseffects}
   \left( \frac{1}{T} \, \vect{q_R} + \, {\mu_{G_R}^{\eta}} \vect{h}_{G_R} \right)  \cdot \Gradient{\Temperature} 
   +
   \,  \vect{h}_{G_R} \cdot \Gradient{ \mu_{G_R} } 
    \le 0 
  \\ &&
  \label{eq:bulk:affinity}
    A_R^{\eqref{eq:actin_polymerization}} \,  w_R^{(\ref{eq:actin_polymerization})}  
    \le 0 
\end{eqnarray} 
\label{eq:bulk:TDrestrictions}
\end{subequations}

\subsubsection{ Decompositions.}

The stress filed $\tensor{ S }$ will be additively decomposed in the sum of the active and passive contributions, analogously to generalized Maxwell models
\begin{equation}
\label{eq:bulk:stressdecap}
\tensor{ S } = \tensor{ S }_{active} + \tensor{ S }_{passive}
\; .
\end{equation}
Active response is related to cytoskeletal reorganization in stress fibers and pseudopodia, whereas the passive response reflects the mechanical behavior that each component of the cell has inasmuch material bodies.

\bigskip
We base the theory for pseudopodia on a multiplicative decomposition of the deformation gradient
\begin{equation}
\label{eq:pseudo:multdecF}
	\tensor{F} = \, \tensor{F}^e \, \tensor{F}^c
\; .
\end{equation}
Tensor $\tensor{F}^c$, named {\em{swelling distortion}} is the local distortion of the material neighborhood of a point due to a volumetric swelling (de-swelling) due to the phase change of actin, from monomeric to a network of filaments and vice-versa. Its representation will be taken as $ \tensor{F}^c = \lambda^c \, \identity$, assuming therefore that a dense network of actin filaments form in pseudopodia. This approach conforms well for lamellipodia filament networks, although it might result inappropriate for slender and highly oriented microstructures seen in filopodia, which might be better captured by the protrusion-contraction uniaxial tensors presented in \cite{Allena:2013aa} or \cite{Hervas-Raluy:2019aa}.
The following identities can be easily assessed:
\begin{equation}
\label{eq:pseudo:Jcidentities}
	\determinant{ \tensor{F}^c } = J^c = {\lambda^c}^3 
	\; ,
	\qquad
	{\dot J}^c / J^c = 3 \,  {\dot \lambda}^c /  {\lambda^c}
	\; ,
	\qquad
	\tensor{l}^c = {\dot{\tensor{F}}}^c \, {\tensor{F}^c}^{-1} = {\dot J}^c / (3 J^c ) \identity
\; .
\end{equation}
We assume that changes in $J^c$ occur because of changes in filaments
$
 J^c = J^c(c_{F_R})
$
and define the partial molar volume of the pseudopodia as 
\begin{equation}
\label{eq:pseudo:pmv}
\Omega_C(c_{F_R}) = \frac{ {\rm d }  J^c}{ {\rm d } c_{F_R} }
\end{equation}
and it holds
\begin{equation}
\label{eq:pseudo:Jdotc}
{\dot J}^c
 =
 \Omega_C(c_{F_R}) 
 \,
 \frac{ {\partial } c_{F_R} }{{\partial }  t}
 \; .
\end{equation}
The decomposition \eqref{eq:pseudo:multdecF}  leads to a multiplicative decomposition for the left Cauchy-Green tensor, too:
\begin{equation}
\label{eq:pseudo:multdecC}
	\tensor{C} = \, \tensor{C}^e \, \tensor{C}^c
\; 
\end{equation}
with the swelling factor $ \tensor{C}^c  = {J^c}^{2/3} \; \mathds{1} $ and the elastic factor 
$
   \tensor{C}^e  = {J^c} ^{-2/3} \; \tensor{C} 
   \; .
$
A classical \cite{AnandJMPS2012} specification of $J^c(c_{F_R})$ is the affine map
\begin{equation}
\label{eq:bulk:Jc_specificaton}
J^c(c_{F_R}) = 1 + (  c_{F_R} - c_{F_R}^0 ) \, \Omega_C
\end{equation}
with a constant partial molar volume $\Omega_C > 0$.

\bigskip
In the realm of viscoelasticity, it is also common to perform a multiplicative decomposition of the deformation gradient $\tensor{F}^e$ into volumetric $\tensor{F}^{e^v}$  and isochoric $\tensor{F}^{e^i}$ factors
\begin{equation}
\label{eq:multdecF}
	\tensor{F}^e = \, \tensor{F}^{e^v} \, \tensor{F}^{e^i}
\; .
\end{equation}
The volumetric factor $ \tensor{F}^{e^v}  = {J^e} ^{1/3} \; \mathds{1} $ turns out to be completely identified by the determinant of $\tensor{F}^e$, 
whereas the isochoric factor 
$
   \tensor{F}^{e^i}  = {J^e} ^{-1/3} \; \tensor{F}^e 
$
obeys to the constraint $\determinant{ \tensor{F}^{e^i} } =1 $. The decomposition \eqref{eq:multdecF}  leads to a multiplicative decomposition for the left Cauchy-Green tensor, too:
\begin{equation}
\label{eq:multdecC}
	\tensor{C}^e = \, \tensor{C}^{e^v} \, \tensor{C}^{e^i}
\; ,
\end{equation}
with volumetric factor $ \tensor{C}^{e^v}  = {J^e} ^{2/3} \; \mathds{1} $ and the isochoric factor 
$
   \tensor{C}^{e^i}  = {J^e} ^{-2/3} \; \tensor{C}^e 
   \; .
$

\subsubsection{Constitutive theory}
\label{sec:constitutivetheory}

%

Two among the several ways to satisfy the thermodynamic restriction (\ref{eq:bulk:TDcrosseffects}) have been discussed in \cite{SalvadoriEtAlJMPS2018} in the framework of trapping. 
Here, we proceed as for the membrane imposing that the cytosol stands in thermal equilibrium, whereby $\Gradient{  \Temperature } = \vect{0}$.
The flow of actin monomers is linearly related to the gradient of their chemical potential by Fick's assumption, consistently with the thermodynamic restriction (\ref{eq:bulk:TDcrosseffects}):
\begin{subequations}
\begin{eqnarray}
\label{eq:bulk:Ficksalpha}
	\vect{h}_{G_R} = - \tensor{M}_{G_R}(c_{G_R}) \; \Gradient{  {\mu}_{G_R} }  
         \; .
\end{eqnarray}
\end{subequations}
The following isotropic non linear specialization for the mobility tensor $\tensor{M}_{G_R} $ is chosen \cite{AnandJMPS2012}
\begin{equation}
\label{eq:bulk:isotropicmobility1}
    \tensor{M}_{G_R} ( c_{G_R} ) = \mobility_{G_R} \, c_{G_R}^{max} \; \vartheta_{G_R} \, \left( 1 -  \vartheta_{G_R} \right)\; \mathds{1} \; ,
\end{equation}
where $c_{G_R}^{max} $ is the saturation limit for receptors, and $\mobility_{G_R}>0 $  is
the {\it{mobility}} of actin monomers.   Assuming that the trapped species $F$ has vanishing mobility is an alternative view of modeling the absence of their flux.

\bigskip
The Helmholtz free energy density $\psi_R$ is modeled by decomposing it into separate parts: a thermal contribution $\psi_R^{th}$,  a diffusive contribution $\psi_R^{diff}$, an elastic contribution $\psi_R^{el}$, and an inelastic (also called {\em{configurational}} ) counterpart $\psi_R^{in}$
\begin{equation}
\label{eq:bulk:psi}
 \psi_R ( \Temperature, c_{G_R}, c_{F_R}, {\tensor {C}},  \tensor{\xi} ) = 
 \psi_R^{th}(\Temperature ) + 
 \psi_R^{diff}( c_{G_R}, c_{F_R} ) + 
 \psi_R^{el}( c_{F_R}, {\tensor {C}} ) + 
 \psi_R^{in}( c_{F_R}, {\tensor {E}},  \tensor{\xi}  ) 
\; .
\end{equation}
This splitting is here taken for granted without motivation. We will not indulge in the description of  $\psi_R^{th}$ (see \cite{SalvadoriEtAlJMPS2018} in case of interest) and we'll rather focus on the remaining parts.

\bigskip
Statistical mechanics depicts the entropy for isolated systems in terms of the density of states, the number of possible molecular configurations \cite{ShellBook2015}.  Making recourse to Stirling's approximation and
since the entropy transforms with the volume by means of $J$,
one finds that the following well-known expression of the entropy of mixing {\em{in the reference configuration}} arises:
\begin{equation}
\label{eq:bulk:etaL_referential}
 \eta_{\beta_R}^{diff} =  - R \, J \, c_{\beta}^{max} \, 
        \left(  
        \vartheta_{\beta} \, \ln[  \vartheta_{\beta}  ] + (1-\vartheta_{\beta}) \, \ln[ 1-\vartheta_{\beta} ]
        \right)  
 \; ,
\end{equation}
the universal gas constant $R$ being the product of Boltzmann constant $k_B$ and Avogadro's number and having denoted with $\beta=G,F$ and with $\vartheta_{\beta_R}$ the ratio 
\begin{equation}
\label{eq:bulk:varthetaRmax}
\vartheta_{\beta_R}( {\vect{X}}, t ) =  {c_{\beta_R}}/{c_{\beta_R}^{max}}
\; .
\end{equation} 
We argued in eq. \eqref{eq:caRmax} that, in view of the structure of the lipid membranes, the maximum number of moles or molecules per unit area for any species remains unchanged in time in the current configuration. The same argument does not seem to apply for the bulk, hence we take henceforth that 
\begin{equation}
\label{eq:bulk:caRmax}
{c_{\beta_R}^{max}}( {\vect{X}}, t)
=
\, c_\beta^{max}({\vect{x}}({\vect{X}}, t), t)  \,  J({\vect{X}}, t)
\end{equation}
is constant and write the free energy density for the continuum approximation of mixing \cite{ShellBook2015} as 
\begin{align}
\label{eq:bulk:psi_diff}
\psi_R^{diff}( c_{G_R}, c_{F_R} ) 
 &
 =
 \;
 \mu_{G_R}^{0} \, c_{G_R} + R \, T c_{G_R}^{max} \left[ \vartheta_{G_R} \ln \vartheta_{G_R} + (1- \vartheta_{G_R} ) \ln (1- \vartheta_{G_R} )\right] 
\\ & \nonumber \;
+
 \mu_{F_R}^{0} \, c_{F_R} + R \, T c_{F_R}^{max} \left[ \vartheta_{F_R} \ln \vartheta_{F_R} + (1- \vartheta_{F_R} ) \ln (1- \vartheta_{F_R} )\right] 
\; .
\end{align}
Note that if the saturation is constant in the current configuration, an explicit coupling of the free energy of mixing with the deformation arises by means of $J$. A new stress would come out, in view of the thermodynamic prescription \eqref{eq:bulk:TDequalities}.

\bigskip
Following \cite{HolzapfelBook}, we will define visco-elastic materials based on the multiplicative decomposition \eqref{eq:multdecC}. Specifically, the free energy for visco-elastic materials will be defined as follows 
\begin{equation}
\label{eq:psi_split}
 \psi_R^{el}( c_{F_R}, {\tensor {C}} ) + 
 \psi_R^{in}( c_{F_R}, {\tensor {E}},  \tensor{\xi}  ) 
 =
  \psi_R^{el, vol}( c_{F_R}, \tensor{C}^{e^v} ) + 
  \psi_R^{el, iso}( c_{F_R}, \tensor{C}^{e^i} ) + 
  \psi_R^{in}( c_{F_R},  \tensor {E}^e -  \tensor{\xi}  ) 
\; .
\end{equation}
with $\psi_R^{in}$ depending upon $ {\tensor {E}}^e$ by means of $ {\tensor {C}^{e^i}} $.
The volumetric part of the elastic free energy is defined through $J^e$, %
highlighting the role of the swelling tensor and of the concentration of pseudopodia, since
\begin{equation}
\label{eq:bulk:Cev}
\tensor{C}^{e^v}  = {J^e} ^{2/3} \; \mathds{1} = {J}^{2/3} \; {J^c}^{-2/3} \; \mathds{1}  =  \left( \frac{ J }{ 1 + (  c_{F_R} - c_{F_R}^0 ) \, \Omega_C}  \right)^{2/3} \; \mathds{1} 
\end{equation}
in view of eq. \eqref{eq:bulk:Jc_specificaton}. On the other end, it holds
\begin{equation}
\label{eq:bulk:Cei}
\tensor{C}^{e^i}  = \tensor{C}^{e} \,  {J^e} ^{-2/3}  = \tensor{C} \,  {J^c} ^{-2/3} \,  {J^e} ^{-2/3}  = \tensor{C} \,  {J} ^{-2/3}  
\end{equation}
hence $\tensor{C}^{e^i}$ depends merely upon the state of deformation and not upon the concentration of species. This outcome reverberates upon the energetic contributions 
$  \psi_R^{el, iso}$ and $  \psi_R^{in}  $.
The latter 
is 
such that 
\begin{equation}
\label{eq:bulk:psi_in_d}
  \frac{ \partial \psi_R^{in} }{ \partial \tensor {E}}  
  = 
  - 
  \frac{ \partial \psi_R^{in} }{ \partial  \tensor{\xi}  }  
   \; ,
\end{equation}
a property physically grounded in the rheological model of Maxwell, for which we refer to \cite{HolzapfelBook} or \cite{SIMOHUGHES}.

\bigskip
Provided that the above holds, the selection for $ \psi_R^{el}$ and $\psi_R^{in}$ is arbitrary. Their selection shall be different in modeling the passive behavior or the active response of pseudopodia and stress fibers.
The elastic, reversible behavior that occurs once the viscous effects vanish (ideally at $t \rightarrow \infty$ ) is captured by $ \psi_R^{el}$. The inelastic free energy accounts for the non-equilibrium response due to viscosity - the so called {\em{dissipation potential}}. By thermodynamic restrictions \eqref{eq:bulk:TDConsistency} and identity \eqref{eq:bulk:psi_in_d}
\begin{subequations}
\begin{align}
\label{eq:internalforces2}
\tensor{\chi}_R &= -  \frac{ \partial \psi_R^{in} }{ \partial  \tensor{\xi}  }  = \frac{ \partial \psi_R^{in} }{ \partial \tensor {E}}  
\\
\label{eq:internalforces3}
\tensor{ S}  &=   2  \, \frac{\partial \psi_R^{el}}{ \partial \tensor {C} }  + \tensor{\chi}_R 
\; .
\end{align}
\end{subequations}
According to eq. \eqref{eq:internalforces3}, tensorial internal forces $\tensor{\chi}_R $ can be interpreted as a {\em{non-equilibrium stress tensor}} of second Piola-Kirchoff kind, that accounts for the viscous response.

\bigskip
Inelastic internal entropy production \eqref{eq:bulk:TDindelsticdiss} was described by the internal flux variables $\tensor{\xi}$ and by their energy-conjugate forces $\tensor{\chi}_R$. A simple way to satisfy constraint \eqref{eq:bulk:TDindelsticdiss} is choosing a positive definite operator $\fourthorder{L}$ such that 
\begin{align}
\label{eq:constinternalforces}
\tensor{\chi}_R &=  \fourthorder{L} \, \dot{ \tensor{\xi}  }  
\; .
\end{align}
In case of isotropy, the fourth order operator $\fourthorder{L}$ restricts to the scalar viscosity $\nu$ times the identity operator. Equations  \eqref{eq:internalforces2}, \eqref{eq:constinternalforces} provide evolution equations for $\tensor{\chi}_R$ that allow the algorithmic integration of the constitutive law once a selection for the free energy densities $\psi_R^{el}$ and $\psi_R^{in}$ is made.

\bigskip

The chemical potential of G-actin monomers and of F-actin networks descends from thermodynamic prescriptions \eqref{eq:bulk:TDequalities}, in the form
\begin{subequations}
\begin{align}
    \mu_{G_R} & = \frac{\partial  \psi_R^{diff}( c_{G_R}, c_{F_R} )  }{\partial c_{G_R}  } 
   \\
    \mu_{F_R} & =
     \frac{\partial \psi_R^{diff}( c_{G_R}, c_{F_R} )   }{\partial c_{F_R} } + 
      \frac{\partial \psi_R^{el, vol}( c_{F_R}, \tensor{C}^{e^v} ) }{\partial c_{F_R} }  +
      \frac{\partial \psi_R^{el, iso}( c_{F_R}, \tensor{C}^{e^i} )  }{\partial c_{F_R} }  +
      \frac{\partial \psi_R^{in}( c_{F_R},  \tensor {E}^e -  \tensor{\xi}  )  }{\partial c_{F_R} }  
\; .				 				 
\label{eq:bulk:chempot}
\end{align}
\end{subequations}
While the chemical potential of actin monomers has merely an entropic nature, mechanical contributions enter the definition of the chemical potential of actin networks.
Specifically, mechanics affects $\mu_{F_R}$ in the volumetric contribution $\psi_R^{el, vol}$  through the swelling tensor $\tensor{C}^{e^v}$ \eqref{eq:bulk:Cev}, whereas the isochoric tensor $\tensor{C}^{e^i}$ was proven to be independent upon the concentration of species in eq. \eqref{eq:bulk:Cei}. Nonetheless, the parameters of the viscoelastic loading-unloading law are expected to depend upon the extent of the polymerization reaction by means of the network concentration $c_{F_R}$ in all terms of the mechanical free energy. 

The mechanical effect on the chemical potential does not propagate into the mass flux because the mobility of actin network is assumed to be negligible. Mechanics however enters the affinity of polymerization reaction \eqref{eq:actin_polymerization} in view of definition  \eqref{eq:bulk:aff}. The stress state is expected to favor polymerization nearby the lipid membrane and depolymerization towards the nucleus.

\subsubsection{The multiscale scenario of cell viscoelasticity }

Although the mechanical framework of the free energy depicted above is rather clear, a specialization of the constitutive equations has not been attempted here and in many cases (as for stress fibers and microtubules) it has not been attempted in the literature, to the best of our knowledge. The complexity leads in the multiscale scenario of cell viscoelasticity:
while the mechanical behavior and properties of intermediate filaments, actin filaments, and microtubules has been nowadays quite clarified, at least in terms of relative stiffness and strengths, bundles of the filaments, their response, polymerization, shape and time evolution are not yet captured by comprehensive models at the ``macroscopic''  scale through appropriate free energies. As a consequence, the ability of models to capture the mechanics of fundamental cellular processes (as chemotaxis, cell sprouting, junction and differentiation, endocytosis and exocytosis to cite a few) still requires abundant research before gaining predicting capabilities in simulations.

The cytoskeleton, an interconnected network of regulatory proteins and filamentous biological polymers, undergoes massive reorganization during cell deformation, especially after cell rolling and adhesion \cite{IntroductiontoCellMechanicsandMechanobiology,WEN2011177} and in mediating, sensing and transduction of mechanical cues from the micro-environment \cite{BARRIGA201955}. 
Homogenized models for the mechanical response of a cell shall include in effective, macroscopic properties the polymerisation/depolimerisation of filaments, the process of cross-linking that determine the architecture of cytoskeletal filaments, and the passive mechanical properties of the cytosol. In view of the above, the thermodynamics of statistically-based continuum theories for polymers with transient networks \cite{BRIGHENTI2017257,VERNEREY20171,VERNEREY2018230,VernereyJMBB2011, LIELEG20094725} appear to be good candidates for the selection of free energies $ \psi_R^{el}( c_{F_R}, {\tensor {C}} ) $
and $ \psi_R^{in}( c_{F_R}, {\tensor {E}},  \tensor{\xi}  ) $. The need of statistical approaches to model the time-dependent response of polymers with reversible cross-links emerges, since the overall response is influenced by rate of assembly and disassembly of cross-linking factors that is controlled at molecular level by actin nucleation, capping, severing factors and by the activity of molecular motors such myosin-II, which, in combination with cross-linkers, appears to be responsible of the viscoelastic properties of the cytoskeleton \cite{MurrelNature2015}. At present however, such a comprehensive model has not yet been proposed for the pseudopodia driven cell motion. Classical models as hyperelastic Saint-Venant \cite{Allena:2013aa} or newtonian viscous fluids \cite{CampbellPF2017} eventually surrounded by a hyperelastic, zero-thickness membrane \cite{Campbell:2020aa} have been used for the pseudopodia, whereas a very large amount of literature concerns pseudopod dynamics ( see for instance \cite{CooperEtAl2012} and the large literature therein ) or ameboid motion \cite{Eidi:2017aa} with no account for their mechanical response. Different approaches to cell motility, as for active gel theory coupled to the classical theory of thin elastic shells, are also widely used \cite{BacherPhysRevE2019}, but are not discussed in this work. The framework described herein, including myosin dynamics, phase transformations between G-actin and F-actin, has been depicted in a set of publications by the group of H. Gomez \cite{Moure:2018aa,Moure:2019aa}. The flow of the F-actin network was treated as a Newtonian fluid and directed by its velocity. A one dimensional yet comprehensive model has been proposed in \cite{PhysRevE.98.062402}.
Not surprisingly, the nucleus and its meshwork of intermediate filaments formed mostly of proteins (nuclear lamina), contribute to the viscoelasticity of cells \cite{ChoJCB2017}. Depending upon the content of Lamins, the nucleus becomes more or less stiff, impacting  cell migration: nuclear deformation facilitates cell migration through complex environments, whereas its stiffness may act as a mechanical barrier for a migratory cell \cite{WolfJCB2013}.  Cells are capable to modify their viscoelasticity while migrating across confined spaces \cite{ThiamNature2016}, a very intriguing mechanism yet complex to be captured macroscopically in view of its multiscale nature. 

The multiscale scenario is invoked also for cell contractility. There are evidences \cite{Fouchard2011} that the interaction among filaments, motors, and cross-linkers is mechanically stimulated. As reported in \cite{BARRIGA201955}, {\em{myosin binding to actin fibers occurs in a force-dependent manner, as well as the contractile response of actomyosin to extracellular stiffness}}. According to \cite{DIZMUNOZ201347}, force feedback controls motor activity and increases density and mechanical efficiency of self-assembling branched actin networks, thus suggesting that those feedbacks could allow migratory cells adjusting their viscoelastic properties to favor migration.
Mass transport and {\em{cell contractility}} have been accounted for in several publications with different degree of complexity \cite{VernereyJMBB2011, VigliottiBMM2016, Hervas-Raluy:2019aa}: to the best of our knowledge, however, 
the force transmission has always been modeled stemming from the similarity between the sarcomeric structure of stress fibers and the actin-myosin interactions in muscle cells. 
In \cite{deshpandeEtAlPRAS2007} a multi-dimensional network of stress fibers was built on the notion of a representative volume element, in which stress fibers can form in any direction with equal probability. An average macroscopic stress is then recovered from the fiber tension, which in turn is generated by the cross-bridging cycles and described by a Hill-like relation \cite{Hill1938PRSB} of viscoelastic nature. Anisotropic stress fibers distributions have been considered in \cite{VernereyJMBB2011}, making use of Von Mises distribution functions at the ``microscale'' coupled to a directional averaging operator. The active contraction is described in terms of the change of fiber length and its rate of change, with a product formula of viscoelastic origin.
Experimental evidences, however, seem to show that such a resemblance might be questionable in the dynamics and mechanics of endothelial cell spreading \cite{Reinhart-King2005} and hence that the predictive capability of this family of models might be poor for this family of cells.

Finally, the {\em{passive response of the cytosol}}, provided mainly by the intermediate filaments attached to the nuclear and plasma membranes, has been modeled by several authors by means of classical models as linear elasticity \cite{VernereyJMBB2011}, the finite strain generalization of Hooke's law \cite{deshpandeEtAlPRAS2007} or a Neo-Hookean potential energy 
\begin{equation}
\label{eq:refHFEnergyNH}
\psi_R^{el} ( \tensor{C}^e ) = \frac{G_0}{2} \, ( I_1 ( \tensor{C}^e ) - 3 )
 \; ,
\qquad
 \psi_R^{in}(  {\tensor {E}}^e,  \tensor{\xi}  )  = \frac{G_0 -  G_\infty}{G_0} \; \psi_R^{el}(  \tensor {E}^e -  \tensor{\xi}  ) 
 \; ,
\end{equation}
where $G_0$ is the initial shear modulus and $G_\infty$ is the shear modulus at the end of the viscous processes.
This classical choice of Helmholtz free energy is associated to efficient integration schemes, depicted in \cite{SIMOHUGHES}.

\section{Concluding remarks}
\label{sec:conclusions}

In this note, a multi-physics framework of protein relocation on the advecting lipid membrane during cells spreading and motion has been put forward. 
It sets the (continuum) thermodynamic background for simulations of receptor recruitment during migration: simulations carried out in \cite{DamioliEtAlSR2017} stem from a simplified form of the framework and described the limiting factors in vascular endothelial growth factor receptors relocation; similarly, we discussed in \cite{SerpelloniEtAl2020} the relocation of integrins on the membrane and their interactions with growth factor receptors; a companion paper \cite{salvadori_in_preparation} deals with the relocation of vascular endothelial growth factor receptors on advecting lipid membrane during endothelial cell adhesion and spreading. Those simulations may have a significant impact in biology and in the pharmacological treatment of cancer, either in view of their predictive nature in virtual experiments, or by clearly identifying the sequence of processes that limit the relocation of targeted proteins during in vitro experiments.

The present work still has significant limitations, yet by illustrating a complex and rigorous scenario it might be a cornerstone to account for several further processes. To cite a major phenomenon that has been insufficiently discussed here, the proteins transport on the membrane  is crucially coupled to the cytoskeleton reorganization, which is related to the motion of integrins on the membrane: the formation of focal adhesion sites is preliminary to stress fibers generation and contractility.  Internalization of complexes is another occurence not included in this work. Further publications, therefore, will be devoted to extend this framework to these and others challenging tasks. 

We also aimed in this paper at recollecting recent publications from several schools on cell mechanics, encasing them in a unified framework, being aware that a comprehensive account of publications is significantly hard in view of the broadness of the literature in the field. We clarified that for some processes, as for contractility and protrusion, either a thermodynamically consistent formulation has not been devised yet or it stems upon simplistic models that do not account for the microstructural evolution of the biopolymers. Even in this fascinating field, the last word is far from being spoken.

\bigskip \noindent
\textbf{ \large Acknowledgements}

\bigskip
This work has been supported by grants from the company {\it{Ferriera Valsabbia}} through
a liberal donation to fund studies in the field of Mechanobiology, and from {\it{Fondazione Berlucchi}} to Mattia Serpelloni. We gratefully acknowledge pleasant scientific discussions with S. Mitola, E. Grillo, and C. Ravelli from the DMMT at the University of Brescia.

\bigskip \noindent
\textbf{ \large References}
  
\bibliographystyle{elsarticle-num}
\bibliography{/Users/albertosalvadori/Bibliography/Bibliography}

\end{document}